%% file: turner_1937_cs.tex
\documentclass[twocolumn, times]{aastex63}

\usepackage{amsmath,verbatim, nicefrac,color, soul}
\usepackage{natbib,graphicx}
\usepackage[caption=false]{subfig}
\newcommand\Hl[1]{\colorbox{yellow}}
\DeclareCaptionFormat{cont}{#1 (cont.)#2#3\par}

\shorttitle{B1937+21 Cyclic Spectroscopy I. Demonstration of Improved Scintillometry}
\shortauthors{J. E. Turner \lowercase{et al}.}

\begin{document}
\title{A Cyclic Spectroscopy Scintillation Study of PSR B1937+21 \\ I. Demonstration of Improved Scintillometry} 
\input{authors}

\input{abstract.tex}

\section{Introduction}
Pulsar scintillation is a valuable tool for studying the small scale structure of the ionized interstellar medium (ISM), ranging from the detection and characterization of sub-AU scale structures \citep{hsa+05}, to probing the origins of extreme scattering events \citep{Coles_2015}, and examining the origins of phase screens in the ISM \citep{ocker}. However, these interactions between the propagating pulsar signal and the ISM also act as a significant source of noise in pulsar timing efforts. In many cases, gaining a better understanding of these ISM structures can lead to better mitigation efforts of the resulting timing noise \citep{Palliyaguru_2015, lentati_2017, turner_cyc}. It is also crucial to understand the frequency dependence of propagation through the ISM, as the effects seen in pulsar scintillation evolve with these frequencies.
\par As a consequence of interactions between pulsar emission and free electrons along the signal's propagation path, the emission will experience multipath propagation and its image will undergo angular broadening, resulting in interference between multiple ray paths at the observer and a broadening of the observed pulsar signal \citep{lorimer_kramer}. The effects of this interference can be seen in pulsar dynamic spectra, which track the intensity of the observed signal over frequency and time. The widths of bright patches in these spectra, known as scintles, provide a means to study the time-evolving behavior of the ISM. The characteristic widths of these scintles in frequency and time are denoted by the scintillation bandwidth, $\Delta \nu_{\rm d}$, and scintillation timescale, $\Delta t_{\rm d}$, respectively, with the subscript $d$ indicating diffractive scintillation. Additionally, these scintles allow one to  estimate the level of scattering and the corresponding arrival time delay a given pulse may experience. Additionally, if the broadened pulse can be deconvolved into the intrinsic pulsar signal and the pulse broadening function (PBF) of the ISM -- the signal imparted by the ISM onto the pulse as a result of their interaction -- fitting this PBF to an exponential decay function provides another way to quantify the level of scattering present in an observation. (Note that multiple previous studies \citep{1937_cs, turner_cyc} have referred to the PBF  as the impulse response function (IRF), although the former is an ensemble average of the latter.) These delays, $\tau_{\rm d}$, are expected to scale as $\nu^{-4.4}$, where $\nu$ is the observing frequency, under the assumptions that the ISM can be accurately characterized by Kolmogorov turbulence and that the smaller length scales of the wavenumber spectrum dominate \citep{Cordes_1998}. Under the assumption of single-screen scattering, these delays are also inversely proportional to $\Delta \nu_{\textrm{d}}$ \citep{Cordes_1998}. Consequently, $\tau_{\rm d}$ can be inferred from $\Delta \nu_{\textrm{d}}$, which is expected to scale with observing frequency as $\nu^{4.4}$ under the same assumptions mentioned above. 
\par Additional information can be gained about the structure of the ISM along a given line of sight (LoS) by examining the 2D Fourier transform of the dynamic spectrum, also known as the secondary spectrum, of an observation, where interference patterns in the dynamic spectrum can give rise to parabolic structures known as scintillation arcs, whose curvature, in conjunction with the pulsar's LoS velocity across the screen, $\textbf{V}_{\textrm{eff}, \perp}$, and distance, can be used to determine the location of the corresponding scattering screen along the LoS \citep{OG_arcs, walker_2004, cordes_2006_refraction, stine_survey}. Structures within these arcs can also be used to gain insight into the structure of the scattering region \citep{hsa+05}. Under the assumption that pulsar scattering occurs at a thin screen, the arc curvatures are expected to scale with frequency as $\nu^{-2}$ \citep{Hill_2003}.

\par While the majority of PTA pulsars have measured scattering delays that are subdominant relative to their overall timing uncertainties \citep{turner_scat}, delays from the most highly scattered pulsars, such as PSRs J1643$-$1224, J1903+0327, and B1937+21, are pertinent for mitigation in pulsar timing efforts. While PTA studies such as \cite{turner_scat} lack the frequency resolution to properly quantify the scattering delays present in these pulsars, recent studies on the aforementioned pulsars (\cite{main_leap} for PSR J1643$-$1224, \cite{geiger_1903} for PSR J1903+0327, and both this work and \cite{main_leap} for PSR B1937+21) have shown that their measured scattering delays are either comparable to or significantly larger than their median time-of-arrival uncertainties at L-band \citep{Agazie_2023_timing}. Additionally, the LoSs through the ISM to these highly-scattered pulsars are often the most interesting to study. However, observations that use pulsar timing array (PTA)-style observing setups, as well as those that use traditional polyphase filterbanks and incoherent dedispersion, typically lack the frequency resolution necessary to see scintillation structures within their dynamic spectra, particularly at lower observing frequencies. This often leaves observers unable to effectively quantify the level of scattering present or resolve scintillation arcs. However, cyclic spectroscopy is a technique that takes advantage of the periodic and amplitude-modulated nature of the pulsar signal to drastically improve the resolution of baseband observations. Cyclic spectroscopy, used for years within various engineering communities, was first introduced to pulsar timing in \cite{cyc_spec}, where the technique made use of the phase information in the voltage data and exploited the fact that the scintillation time spans many pulse periods, allowing the PBF to be estimated. The cyclic spectrum is defined as 
\begin{equation}
\label{act_cyc}
    {S_E}(\nu, \alpha_k) = \langle E(\nu + \alpha_k/2)E^{*}(\nu - \alpha_k/2)\rangle,
\end{equation}
where $E(\nu)$ is the electric field of the signal and $\alpha_k = k/P$ is the cyclic frequency at the $k^{\textrm{th}}$ harmonic for pulse period $P$ \citep{dsj+20}. To account for changes in the pulsar signal due to the ISM, over integration times similar to a pulsar's scintillation timescale, we can write the cyclic spectrum as 
\begin{equation}
\label{prac_cyc}
    {S_E}(\nu, \alpha_k) = \langle H(\nu + \alpha_k/2)H^{*}(\nu - \alpha_k/2)\rangle{S_x}(\nu, \alpha_k),
\end{equation}
where $H(\nu)$ is the transfer function of the ISM, which is the Fourier transform of the ISM PBF, and ${S_x}(\nu, \alpha_k)$ is the Fourier transform of the intrinsic pulse profile \citep{dsj+20}. 
\par As a result of the additional information made available in the pulsar signal through this technique, while maintaining a given number of pulse phase bins, one can acquire significantly higher frequency resolution than accessible through the use of traditional polyphase filterbank channels, allowing for the emergence of previously unresolved features in periodic and dynamic spectra while maintaining high pulse phase resolution. This resolution improvement has already been used to great effect in millisecond pulsar (MSP) observations at low radio frequencies \citep{1937_cs, Archibald_2014} and could lead to pulsar timing procedures where scintles and scintillation arcs (assuming observing durations where scintillation timescales are obtainable) are resolved in all epochs. This would allow for scintillation arcs to be studied with preexisting monthly observations of many pulsars without the need to alter the observing parameters crucial to PTA feasibility, and providing many LoSs through the ISM over which to examine arc evolution over frequency and time \citep{main_leap}. In particular, these high cadence observations are crucial for tracking arc curvature over time, allowing for tighter constraints on scattering screen distances.
\par Additionally, this scintle resolution would make it possible to estimate scattering delays at all epochs, provided flux density is sufficient for visible scintles in dynamic spectra. This would (perhaps dramatically) advance the feasibility of correcting for scattering delays in the pulsar timing models used by PTAs. Arguably, the main benefit of CS is its ability to resolve PBFs, providing a direct approach to mitigating scattering delays across individual scintillation timescales. For this reason, there are ongoing efforts to implement a CS backend at Green Bank Observatory, which would allow for real-time CS processing on MSPs. The resulting data could then be used to recover a pulsar's PBF and subsequently measure and correct the related scattering delay. With the inclusion and automation of this scattering delay correction pipeline, we would not use the scattering delays indirectly estimated through dynamic spectra, and these higher-resolution spectra would instead primarily be used to place better constraints on the ISM. However, the capability of high-resolution dynamic spectra also allows for the monitoring of scattering delays in highly scattered pulsars, potentially informing red noise modeling efforts to remove these effects in instances when we cannot recover PBFs. While PTAs monitor these delays over time \citep{shapiroalbert2019analysis,Levin_Scat,turner_scat,epta1}, they do not account for them in timing models. This results in a noise floor of tens to hundreds of nanoseconds at L-band, depending on the pulsar, and even higher at lower observing frequencies \citep{turner_scat}. 
\par There are attempts within PTAs to indirectly account for the effects of scattering. For example, some PTAs fit $\nu^{-4}$ delays to TOAs \citep{epta_noise}, but this approach ignores that scaling indices have been observed to vary across both LoSs and epochs. Ignoring this variability could result in improper correction of these scattering effects, thus lowering S/N and reducing  sensitivity to gravitational wave signals in PTA data \citep{turner_scat}. For highly scattered pulsars, which can have many scintles across an observing band of only a few hundred MHz, the frequency resolution improvement provided by CS could allow for more accurate, frequency-dependent scaling law measurements on individual epochs. In instances where full PBF deconvolution is not possible, this improved accuracy could potentially inform PTA noise modeling by providing priors to red noise analyses that attempt to mitigate delays from interstellar scattering. 
\par The ideal situation when performing frequency-dependent scintillation analyses of pulsar emission propagation through the ISM is to obtain all observations at the various frequencies simultaneously so that the same LoS is being sampled throughout (although, note the small discrepancies in effective column density that can arise due to frequency-dependent scattering volumes; see \citealt{Freq_DM} for details.). Typically when frequency-dependent analyses are performed, they are limited by the frequency resolution and bandwidth of the observations, as well as the observing cadence between observations at different frequencies. Some of these analyses used a large frequency range, but only 2$-$4 measurements \citep{Bhat_2004}, with these measurements sometimes taken on different days \citep{Hill_2003,epta1}. In general, this should not affect results provided they were taken within a pulsar's refractive timescale and the effective velocity remains similar over the course of the observations, but nearly simultaneous measurements are preferred. Other methods make simultaneous measurements across many frequencies \citep{Krishnakumar_2019, turner_arcs}, although all such studies have been performed at lower frequency ranges \citep{Archibald_2014, Bansal_2019}. Studies at L-band have taken the approach of breaking up large bandwidth observations into smaller frequency slices as well, although this appears to have been limited to 200 MHz slices across 800 MHz bands \citep{Levin_Scat,turner_scat}. 
\par In this paper we use cyclic spectroscopy to acquire fine frequency resolution across a relatively small bandwidth, allowing for eight simultaneous measurements per epoch for frequency-dependent analyses. In Section \ref{sec:data} we present the data used for this work. In Section \ref{sec:analyses} we discuss the data processing and analyses performed on these data. Section \ref{sec:results} discusses the results of these analyses and their comparison to theory and previous literature.  Finally, in Section \ref{concl} we discuss our conclusions and the potential for future work that can take full advantage of the cyclic spectroscopy techniques discussed in this work. 
\par While this paper focuses on the benefits of cyclic spectroscopy (CS) in the context of analyses of the ISM in PTA observing setups, as well as providing parameter estimations and scaling law fits of these data, accompanying papers (Dolch et al., in prep, Turner et al., in prep) will focus on a more detailed interpretation of these measurements.
\section{Data}
\label{sec:data}
\begin{figure*}[!ht]
\centering
\hspace*{-1cm}                                                           
\includegraphics[scale = 0.58]{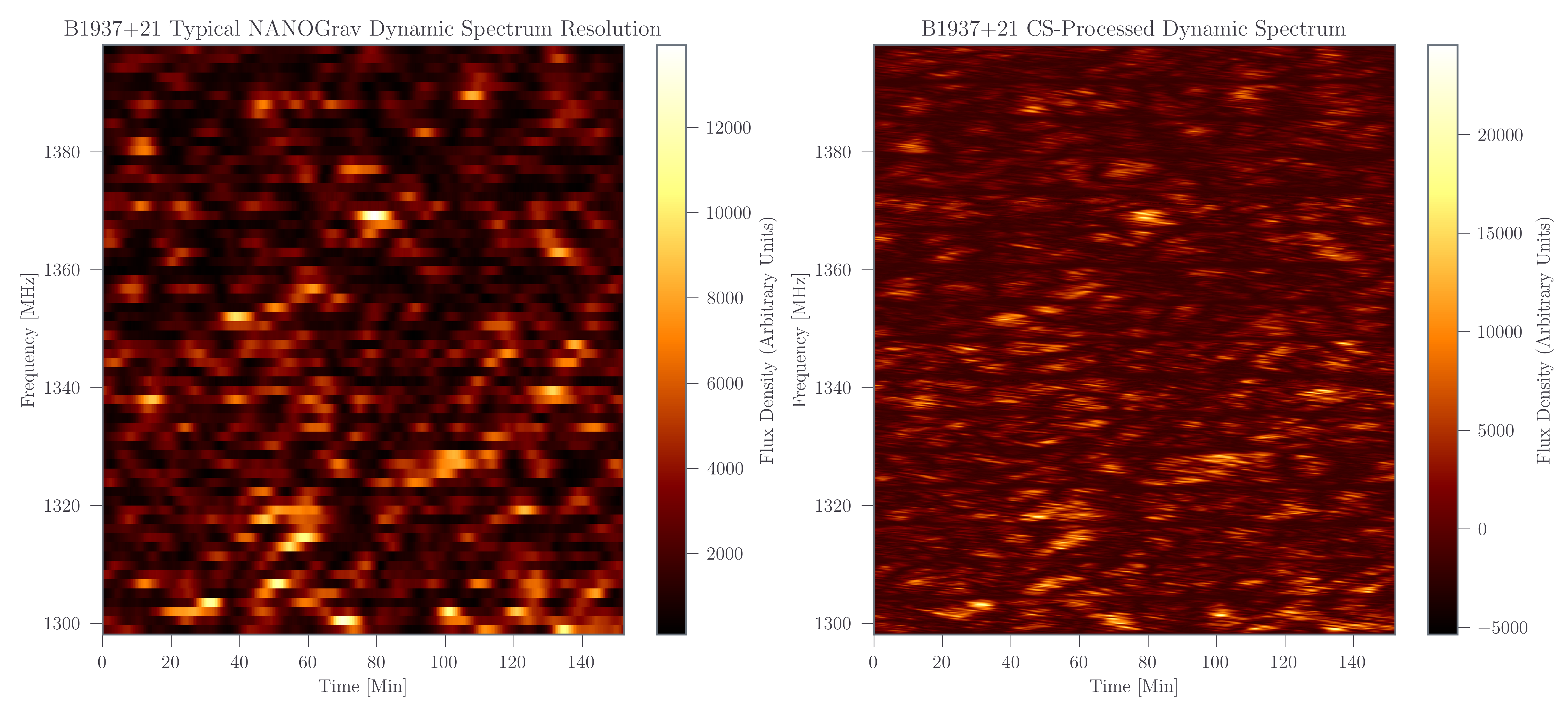}
\caption{A 100 MHz slice of the dynamic spectrum on MJD 56198 showing data processed using Fourier spectroscopy that yields 1.5 MHz resolution with the standard frequency resolution available to NANOGrav (1.5 MHz channels, left) and using cyclic spectroscopy (6.1 kHz channels, right), both using 1024 pulse phase bins. Cyclic spectroscopy greatly improves the frequency resolution of the dynamic spectrum while maintaining high time resolution, allowing for high time and frequency resolution within the same data product. The darker bands in the image on the right are the gaps between the filterbanks.}
\label{ex_cs_compare}
\end{figure*}

Our observations of PSR B1937+21 (P2627 PI Stinebring) took place on MJDs (and fractional years) 56183 (2012.70), 56198 (2012.74), and 56206 (2013.01). Observations on MJDs 56183 and 56206 spanned approximately 2 hours each, while the observation on MJD 56198 spanned approximately 2.5 hours, and all observations were taken in baseband mode with the FPGA-based PUPPI spectrometer at the Arecibo Observatory using 200 MHz of bandwidth with 6.25 MHz wide filterbank channels centered at 1373.125 MHz. Our data were recorded on eight 25 MHz-wide banks, which we maintained for our subband analyses. Our raw baseband data were then processed via cyclic spectroscopy using \texttt{dspsr}\footnote{\url{http://dspsr.sourceforge.net/}} \citep{dspsr} with 1024 pulse phase bins and 1024 cyclic channels per filterbank channel, for a final channel resolution of around 6.1 kHz. We used blocks of data averaged over approximately 28 second intervals for our analyses. An example dynamic spectrum processed from these baseband data using the typical NANOGrav frequency resolution of 1.5 MHz \citep{Agazie_2023_timing} compared with the same dynamic spectrum processed with cyclic spectroscopy is shown in Figure \ref{ex_cs_compare}. 

\section{Analyses}
\label{sec:analyses}
\subsection{Dynamic Spectra}
After processing our baseband data using CS to improve frequency resolution, for each observation, we summed the two orthogonal polarizations and created a dynamic spectrum, defined as
\begin{equation}
\label{dynspec}
S(\nu,t)=\frac{P_{\rm{on}}(\nu,t)-P_{\rm{off}}(\nu,t)}{P_{\rm{bandpass}}(\nu,t)},
\end{equation}
where $S$ is the intensity of the pulsar signal at each observing frequency, $\nu$, and time, $t$, ${P}_{{\rm{bandpass}}}$ is the total power of the observation, and ${P}_{{\rm{on}}}$ and ${P}_{{\rm{off}}}$ are the integrated power in all on- and off-pulse components, respectively, at each frequency and time (see Figure \ref{ex_cs_compare}). The on-pulse region was determined by first creating a template from the average profile of the corresponding subband and then performing an overlap integral of $P(\nu,t)$ with that template, in which we integrated over the product of the profile at a given $\nu$ and $t$ and the template.

\par To examine a given dynamic spectrum's characteristic scintle widths in frequency and time, we acquired its autocorrelation functions (ACFs) in both frequency and time using \textsc{pypulse} \citep{pypulse} to extract that spectrum's 2D ACF. We then took slices at zero time lag and zero frequency lag to acquire time and frequency ACFs, respectively. As a consequence of the high S/N in our data, noise spikes in both the frequency and time ACFs were insignificant enough that no modification to the ACFs was necessary prior to fitting. To determine the scintillation bandwidth for a given subband, we took its dynamic spectrum and retrieved its 1D frequency ACF. We then fit its frequency ACF with a Lorentzian and measured the scintillation bandwidth, $\Delta \nu_{\rm d}$, by finding the half width at half maximum of its fit. Similarly, to determine the scintillation timescale for a given subband, we took its dynamic spectrum and fit its time ACF with a decaying exponential and measured the scintillation timescale, $\Delta t_{\rm d}$, by finding the half width at $1/e$ of the fit \citep{Cordes_1985}. 
\par 

As with all pulsar scintillation studies, our precision in estimating scintillation bandwidths and timescales from ACFs was limited by the number of scintles visible in a given observation. As such, our uncertainties were dominated by the finite scintle effect, and are quantified by 
\begin{equation}
\label{finite_scintle}
\begin{split}
\epsilon &  = \frac{\Delta \nu_{\rm d}}{2\ln(2) N_{\rm scint}^{1/2}} \\
& \approx \frac{\Delta \nu_{\rm{d}}}{2\ln(2)[(1+\eta_{\text{t}}T/\Delta t_\text{d})(1+\eta_\nu B/\Delta \nu_\text{d})]^{1/2}},
\end{split}
\end{equation}
with ${N}_{{\rm{scint}}}$ being the number of scintles observed over the course of the observation, $T$ and $B$ being the total integration time and total bandwidth, respectively, and ${\eta }_{{\rm{t}}}$ and ${\eta}_{\nu}$ being filling factors that can range from $0.1$ to $0.3$ depending on how one defines the characteristic timescale and bandwidth, which for this paper are both set to 0.2 \citep{Cordes1986}. Throughout all of our observations, this resulted in 90$-$160 scintles in the lowest frequency subbands and around 40$-$120 scintles in the highest frequency subbands. The equivalent finite scintle error for scintillation timescale is identical to the scintillation bandwidth error aside from the factor of $\ln(2)$. These uncertainties were then added in quadrature with the uncertainties derived from the covariance matrices of ACF fits to obtain the total uncertainty for a given scintillation parameter estimation. 
\par
\begin{figure}[t]
\centering
\includegraphics[scale = 0.58]{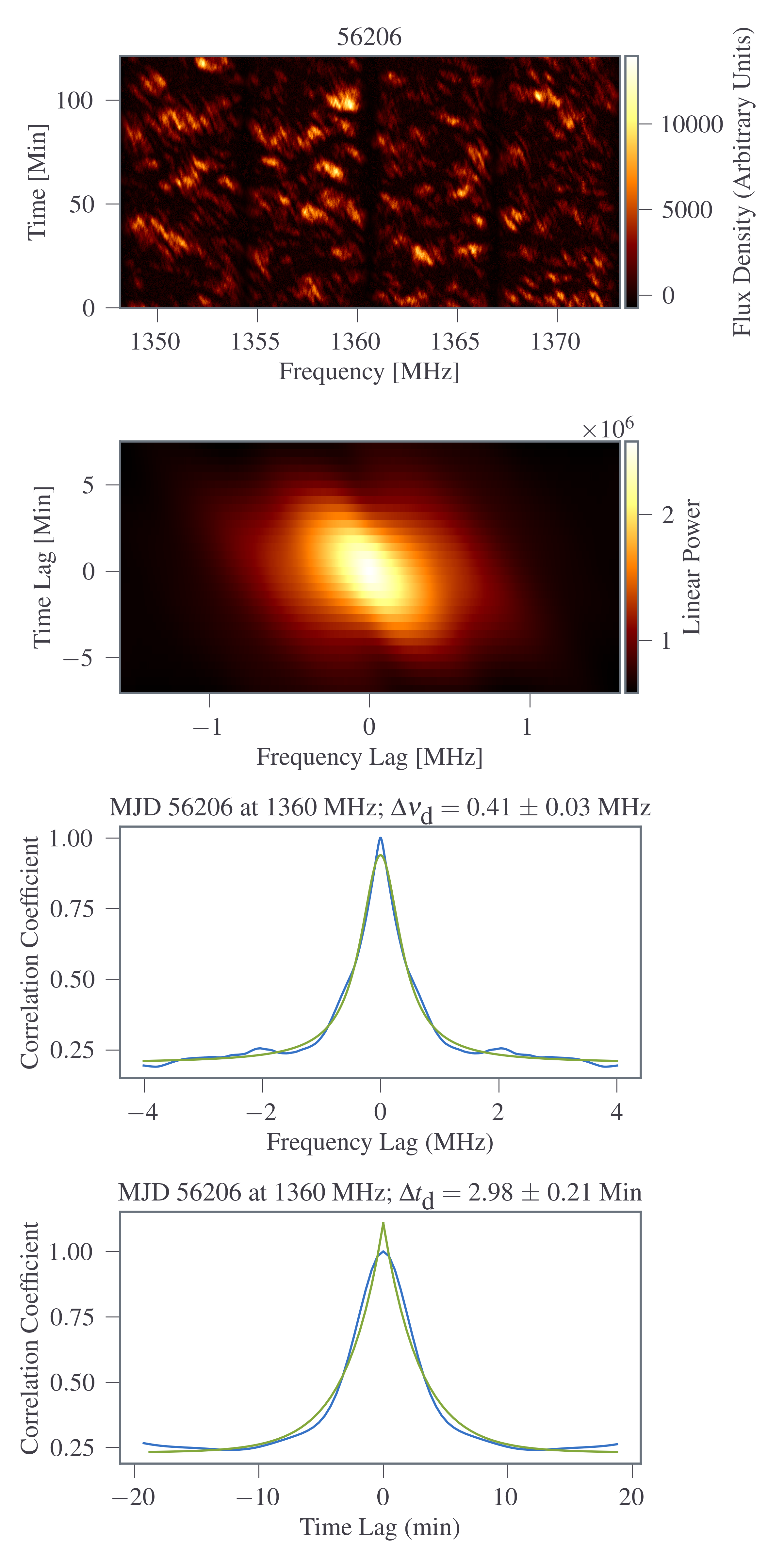}
\caption{(a) A dynamic spectrum from MJD 56206 taken at 1360 MHz. (b) The corresponding 2D autocorrelation function. (c) The 1D frequency slice (blue) of the 2D autocorrelation function with its corresponding Lorentzian fit (green). (d) The 1D time slice of the 2D autocorrelation function (blue) with its corresponding one-sided decaying exponential fit (green). Despite the strong slant in the 2D ACF, due to the low transverse velocity of PSR B1937+21, contributions to scattering due to refraction are negligible in these observations \citep{hewish_scint}.}
\label{ex_acf_fits}
\end{figure}

\begin{figure}[t]
    \centering
    \includegraphics[scale = 0.58]{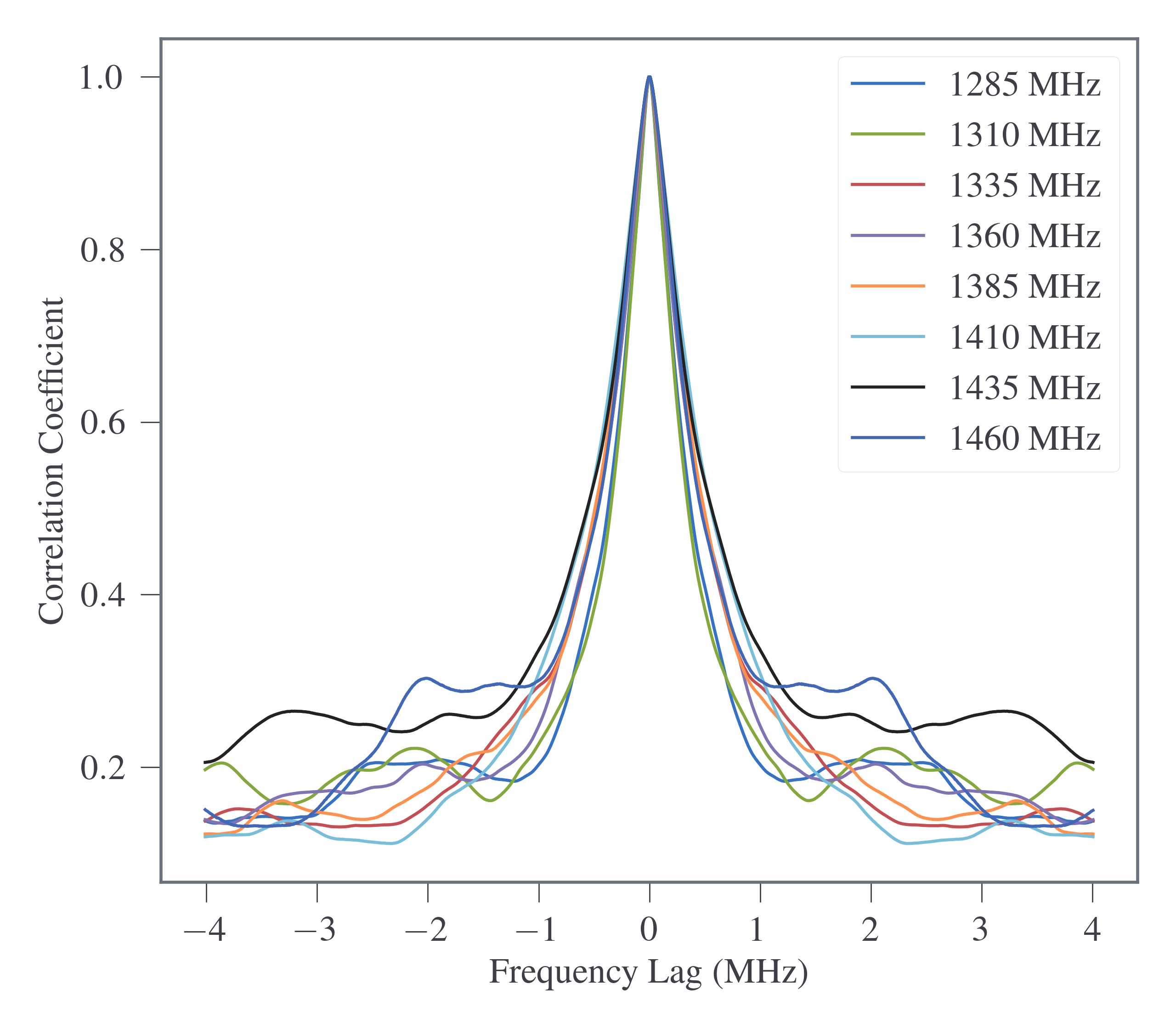}
    \caption{Frequency ACFs for all observing frequencies on MJD 56206}
    \label{acf_set}
\end{figure}

 A complete list of all ACF-measured scintillation parameters can be seen in Table \ref{scint_table}. An example dynamic spectrum, along with its 2D ACF and 1D ACF fits, are shown in Figure \ref{ex_acf_fits}, while an example set of all frequency ACFs from a single epoch is shown in Figure \ref{acf_set}.  Also of note in our observations, and visible in Figure \ref{ex_acf_fits}, is the strong slant in the 2D ACF, indicating a non-zero scintillation drift rate, $\rm{d}\nu/\rm{d}t$. 

\begin{deluxetable}{CCCCC}

\tablewidth{0pt}
\tablecolumns{5}

\tablecaption{Subband Scintillation Bandwidth and Timescale Measurements \label{scint_table}}
\tablehead{ \colhead{MJD} & \colhead{Frequency} & \colhead{$\Delta \nu_{\textrm{d}}$} & \colhead{$\Delta t_{\textrm{d}}$} &  \colhead{$N_{\textrm{scint}}$} \\ \colhead{} & \colhead{\text{(MHz)}} & \colhead{\text{(MHz)}} & \colhead{\text{(min)}} & \colhead{} \vspace{0.05cm}}
\startdata
56183 & 1285 & 0.42 $\pm$ 0.03 & 3.56 $\pm$ 0.24 & 95\\
56183 & 1310 & 0.47 $\pm$ 0.03 & 3.28 $\pm$ 0.25 & 100\\
56183 & 1335 & 0.49 $\pm$ 0.04 & 4.15 $\pm$ 0.29 & 78\\
56183 & 1360 & 0.42 $\pm$ 0.03 & 3.58 $\pm$ 0.24 & 102\\
56183 & 1385 & 0.45 $\pm$ 0.03 & 3.77 $\pm$ 0.30 & 93\\
56183 & 1410 & 0.51 $\pm$ 0.04 & 4.34 $\pm$ 0.30 & 72\\ 
56183 & 1435 & 0.42 $\pm$ 0.03 & 4.04 $\pm$ 0.30 & 92\\
56183 & 1460 & 0.66 $\pm$ 0.07 & 5.40 $\pm$ 0.49 & 48\\
\hline
56198 & 1285 & 0.53 $\pm$ 0.04 & 3.96 $\pm$ 0.24 & 91 \\
56198 & 1310 & 0.69 $\pm$ 0.05 & 2.95 $\pm$ 0.22 & 93 \\
56198 & 1335 & 0.52 $\pm$ 0.03 & 2.88 $\pm$ 0.16 & 122 \\
56198 & 1360 & 0.72 $\pm$ 0.06 & 4.07 $\pm$ 0.29 & 67 \\
56198 & 1385 & 0.62 $\pm$ 0.05 & 3.34 $\pm$ 0.21 & 92 \\
56198 & 1410 & 1.04 $\pm$ 0.11 & 4.50 $\pm$ 0.36 & 45 \\
56198 & 1435 & 1.10 $\pm$ 0.12 & 4.78 $\pm$ 0.47 & 41 \\
56198 & 1460 & 0.81 $\pm$ 0.07 & 3.89 $\pm$ 0.31 & 63 \\
\hline
56206 & 1285 & 0.32 $\pm$ 0.02 & 2.88 $\pm$ 0.19 & 154 \\ 
56206 & 1310 & 0.28 $\pm$ 0.02 & 3.11 $\pm$ 0.20 & 164\\
56206 & 1335 & 0.46 $\pm$ 0.03 & 2.73 $\pm$ 0.21 & 117\\
56206 & 1360 & 0.41 $\pm$ 0.03 & 2.98 $\pm$ 0.21 & 120\\
56206 & 1385 & 0.46 $\pm$ 0.03 & 3.36 $\pm$ 0.23 & 98\\
56206 & 1410 & 0.54 $\pm$ 0.04 & 3.62 $\pm$ 0.26 & 79\\
56206 & 1435 & 0.41 $\pm$ 0.03 & 2.94 $\pm$ 0.20 & 123\\
56206 & 1460 & 0.44 $\pm$ 0.03 & 3.22 $\pm$ 0.22 & 106
\enddata
\tablecomments{Scintillation bandwidth and timescale measurements, as well as the calculated number of scintles, across all subbands and epochs. All errors shown are 1$\sigma$.}
\vspace{-1.05cm}
\end{deluxetable}

\par The scintillation bandwidth and scintillation timescale scaling indices, ${x_{\Delta \nu_{\rm d}}}$ and $x_{\Delta t_{\rm d}}$, at each epoch were determined by performing a weighted linear least-squares fit of the form
\begin{equation}
\label{scattering_index_fit}
\Delta \nu_{\textrm{d}}=a_{\Delta \nu_{\rm d}}\nu^{x_{\Delta \nu_{\rm d}}},
\end{equation}
and 
\begin{equation}
\label{timescale_index_fit}
\Delta t_{\rm d} =      a_{\Delta t_{\rm d}}\nu^{x_{\Delta t_{\rm d}}},
\end{equation}
with the weights being the squared inverses of uncertainties on the scintillation bandwidths and timescales, respectively, across all scintillation bandwidth and timescale measurements from a given MJD. 
\par We do not see any evidence of the filterbank gaps influencing either our ACF analyses using 25 MHz subbands.  However, their effects are present in full-band frequency ACFs. Many peaks emerge at non-zero lag in the full-band frequency ACFs that are a significant fraction of the height of the zero-lag peak. However, efforts to remove the filterbank gaps from the dynamic spectra by setting them equal to the the background noise in the spectra yielded no significant change in our scintillation bandwidth estimations, full-band or otherwise. As a result, we can conclude that the filterbank gaps make no significant difference in our scintillation bandwidth measurements.

Our observations are separated by 15 and 8 days (see Table \ref{scint_table}.) By calculating refractive timescales for each epoch at each frequency and then taking a weighted average, we can check whether our scintillation measurements are independent from each other. From \cite{stinecon}, the refractive timescale, $t_r$, can be approximated by 
\begin{equation}
    t_r\approx\frac{4}{\pi}\Big(\frac{\nu\Delta t_{\rm d}}{\Delta \nu_{\rm d}}\Big).
\end{equation}

\noindent Using the above equation, we find weighted average refractive timescales of 9.9 $\pm$ 0.4, 6.2 $\pm$ 0.2, and 8.9 $\pm$ 0.3 days for epochs 56183, 56198, and 56206 respectively. Based on these values, we can conclude that MJDs 56183 and 56198 do not fall within the same refractive timescale and are consequently independent from each other. MJDs 56198 and 56206 may marginally fall within the same refractive timescale, although the refractive timescale of 6.2 days for MJD 56198 renders this claim inconclusive.
\subsection{Secondary Spectra}
\par
\begin{figure*}[!ht]
\centering
\hspace*{-1cm}                                                           
\includegraphics[scale = 0.58]{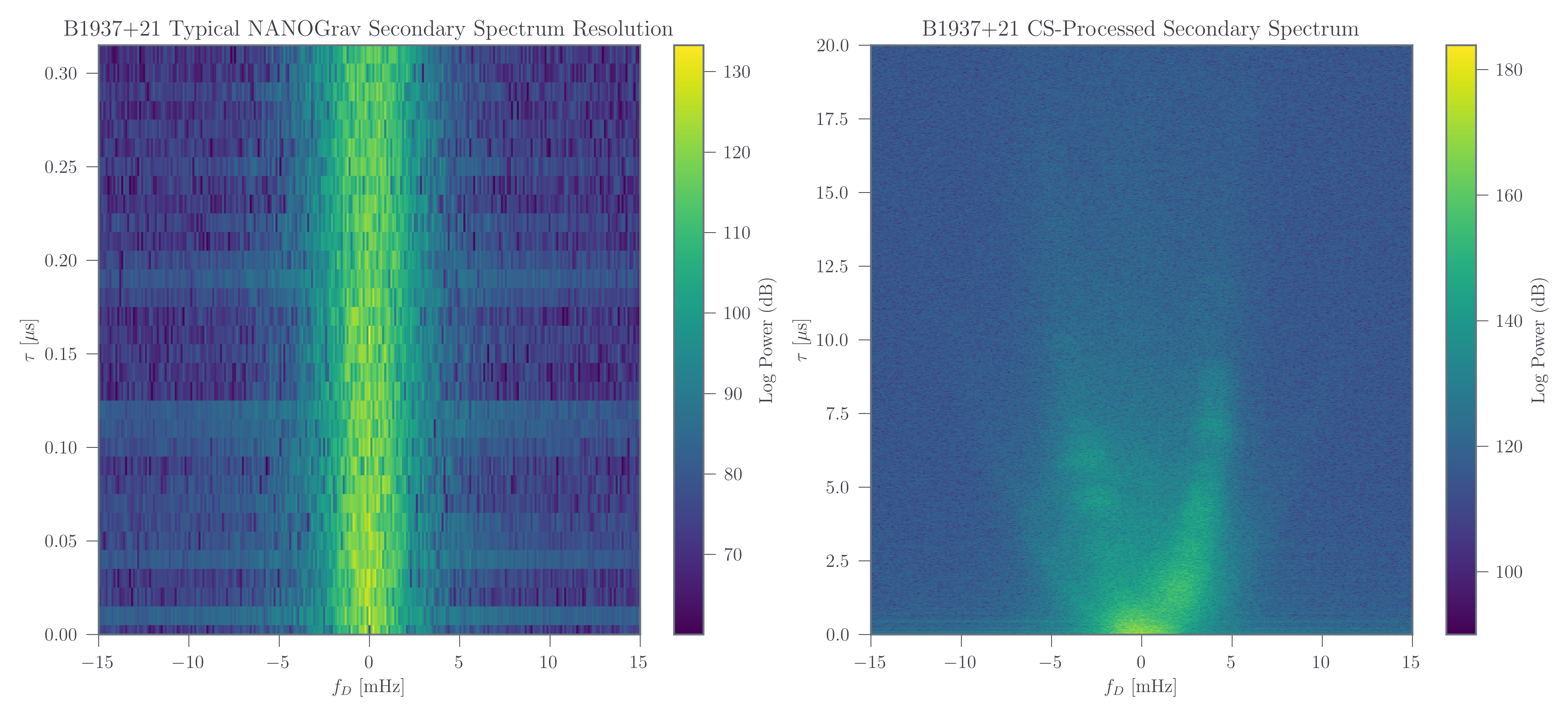}
\caption{Secondary spectra computed from the two dynamic spectra shown Figure \ref{ex_cs_compare}. The two order-of-magnitude difference in the y-axis ranges is a consequence of the different delay Nyquist limits, which are caused by the different frequency resolutions of their corresponding dynamic spectra.}
\label{ss_compare}
\end{figure*}

 To create the secondary spectrum for each observation, we took the squared modulus of the two-dimensional Fourier transform of the corresponding dynamic spectrum (see Figure \ref{ss_compare}). 

\par At each frequency slice, to measure the curvature of each arm in a given arc, we followed the approach described in \cite{stine_survey} whereby we divided each secondary spectrum along the center of its fringe frequency axis and fit the maximum arc value at each delay using $\tau=\eta f_D^2$, where $\tau$ is the differential time delay (Fourier conjugate variable to observing frequency), $\eta$ is the arc curvature, and $f_D$ is the fringe frequency (Fourier conjugate variable to time). We note that, while two arcs appear to be present in all secondary spectra, we only applied fits to the more prominent, higher curvature arc in each spectrum.

\par Following \cite{Hill_2003}, to measure scintillation arc scaling indices, which we refer to as $x_\eta$, we performed a weighted linear least-squares fit of the form
\begin{equation}
\label{arc_scale_fit}
    \eta = a_\eta \nu^{x_\eta},
\end{equation}
with the weights being the squared inverses of the uncertainties on each arc curvature, using the measured curvatures of a given arm at all frequencies within a given MJD. This fit was performed separately for arc measurements on the left and right sides of the fringe frequency axis to acquire scaling indices $x_{\eta,L}$ and $x_{\eta,R}$, respectively.

\section{Results \& Discussion}
\label{sec:results}

\subsection{Scintillation Bandwidth and Timescale Measurements \& Scaling}
\par Although scattering delays are not currently accounted for explicitly in PTA timing models, the ISM introduces time-variable delays on the order of tens to hundreds of nanoseconds at center frequencies around 1.4 GHz in MSPs timed by various PTAs \citep{Levin_Scat,turner_scat,epta1}. Observing with larger frequency bandwidths can improve precision of timing model parameters, such as correcting for the frequency-dependent effects of DM. However, wider bandwidths also mean a wider range of frequencies over which effects of the ISM can vary in influence. Additionally, pulse profiles evolve with frequency and scintillation introduces further frequency-dependent modulations in pulse intensity, adding stochastic noise to this chromatic behavior. In the North American Nanohertz Observatory for Gravitational Waves (NANOGrav) PTA collaboration, current practice corrects for time-variable dispersion measure (DM) using a piece-wise fitting procedure \citep{Agazie_2023_noise}. If a similar approach were to be used to account for scattering delays at a given epoch, we might assume the medium exhibits Kolmogorov turbulence and use a $\nu^{-4.4}$ dependence for scattering delays across the band. However, it has been shown that many LoSs frequently deviate from this dependence and have scaling indices that fluctuate across epochs \citep{turner_scat}. There is also a strong covariance between DM and scattering delay determined this way as a consequence of DM fits absorbing scattering delays \citep{turner_undergrad}. As such, this approach could result in TOA misestimations of tens to hundreds of nanoseconds depending on the pulsar. For this reason, it becomes crucial to understand how scintillation bandwidth, and, consequently, scattering delay, vary across these wide observing bands when attempting to correct for scattering effects. 
\par
\begin{figure*}[!ht]
    \captionsetup[subfigure]{labelformat=empty}
    \subfloat[\centering ]{\hspace*{-.45cm} {\includegraphics[width=0.36\textwidth]{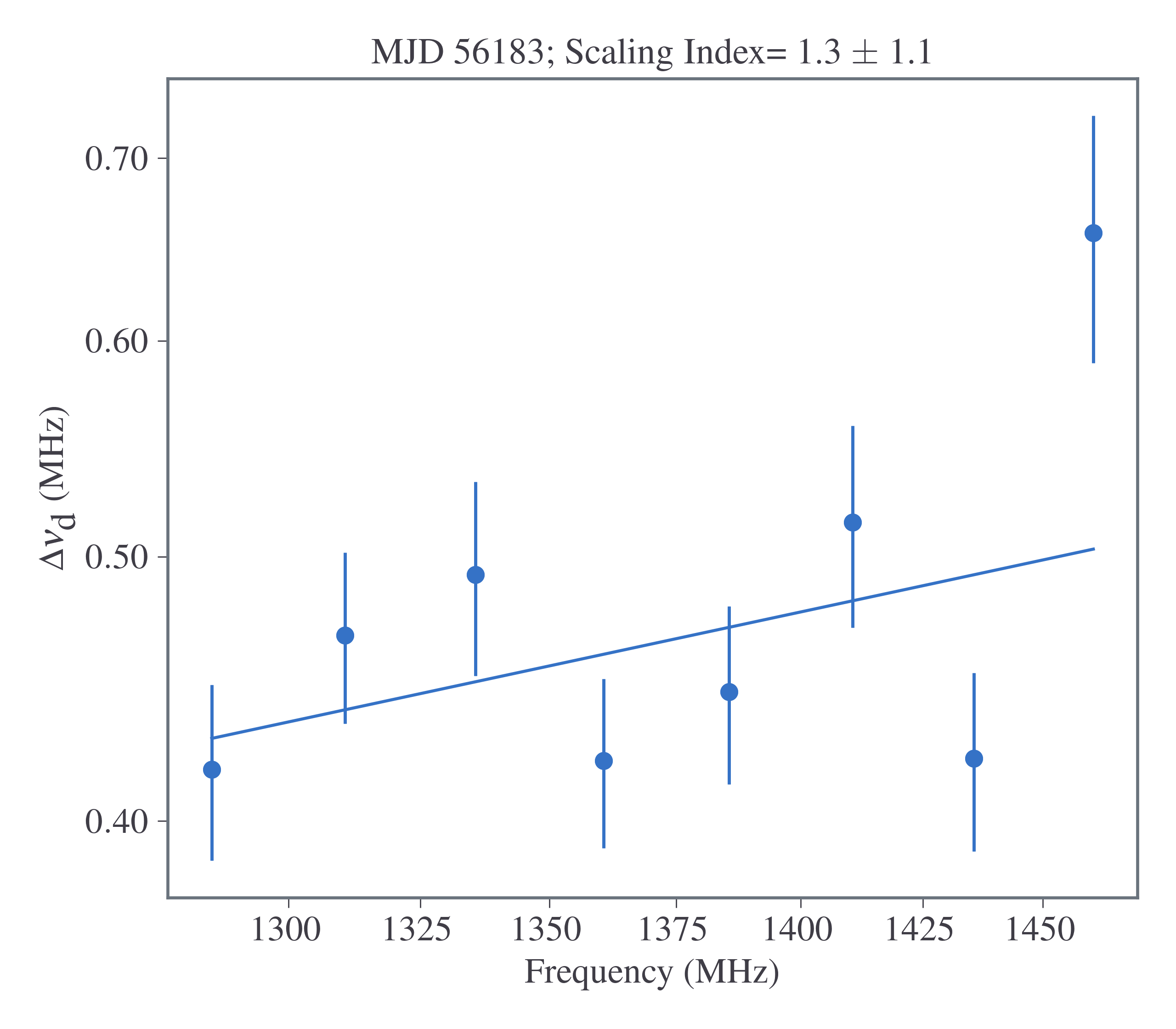} }}%
    \subfloat[\centering]{\hspace*{.0cm} {\includegraphics[width=0.36\textwidth]{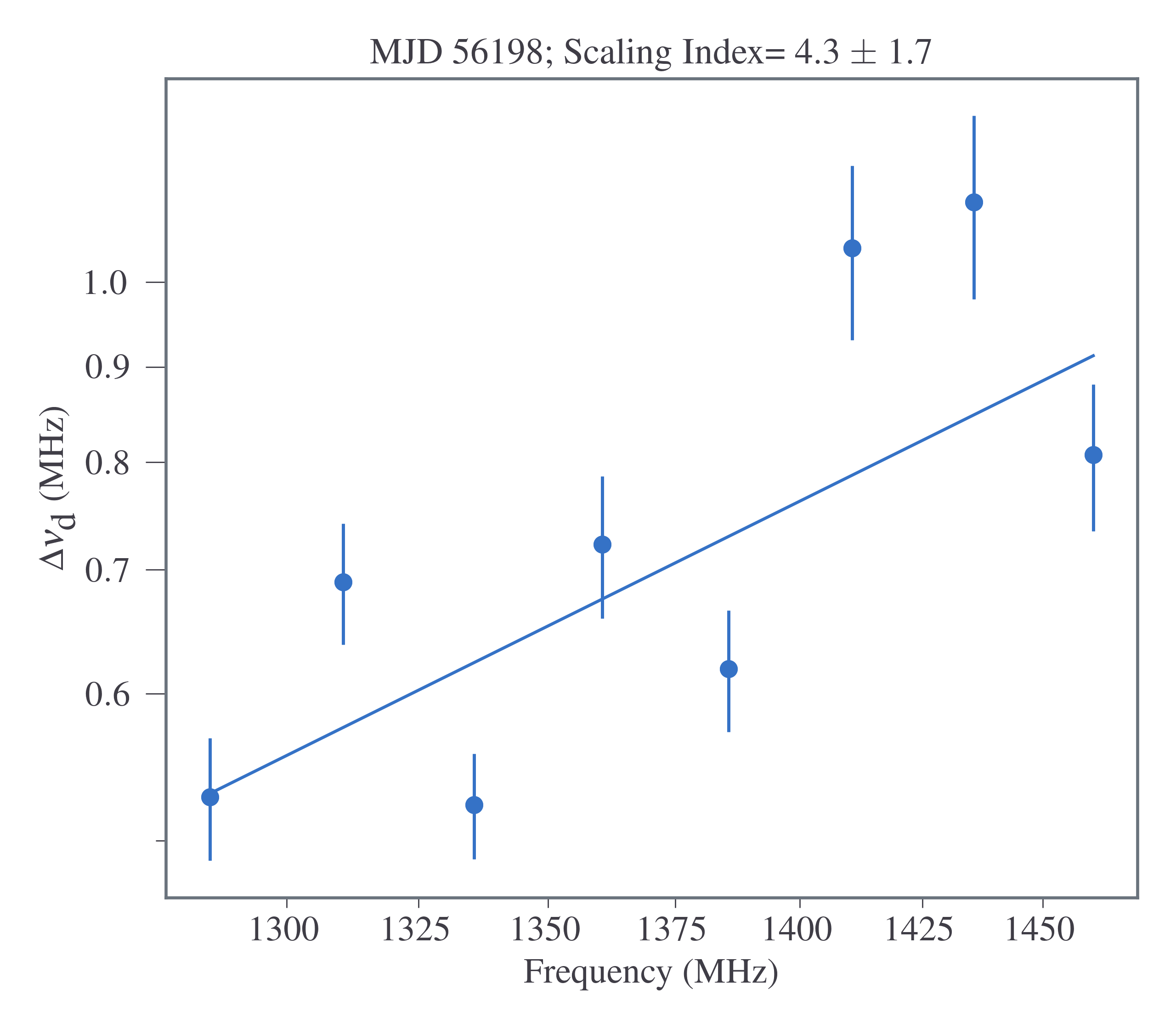} }}%
    \subfloat[\centering]{\hspace*{.1cm}{\includegraphics[width=0.36\textwidth]{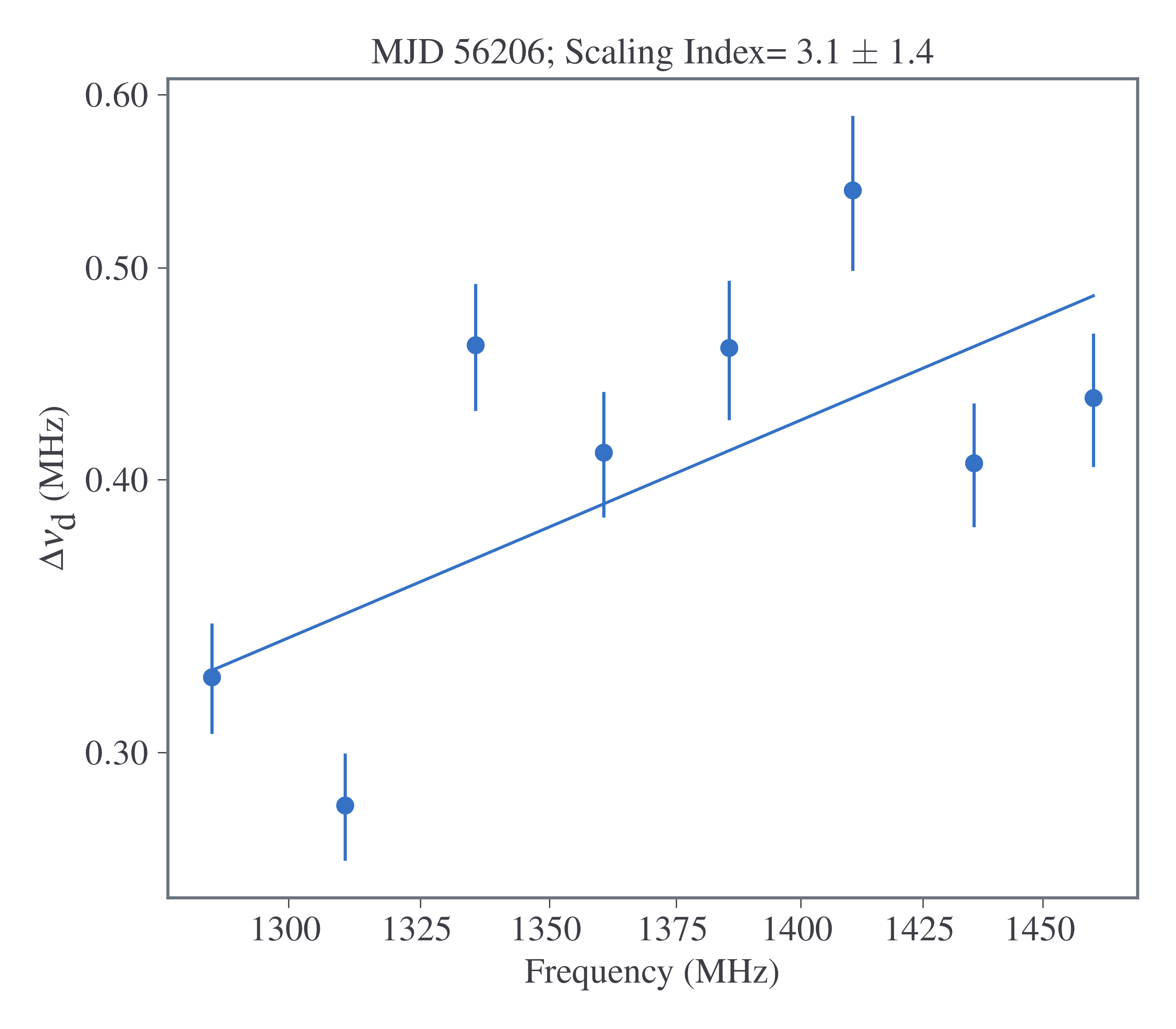} }}%
    \caption{Scintillation bandwidth scaling indices, from left to right, on MJDs 56183, 56198, and 56206, with scaling index fits indicated by the blue lines. 1$\sigma$ error bars are shown. Both axes are in log space on all plots.}%
    \label{ex_scale_fit_delay}%
\end{figure*}

\begin{figure}[t]
    \centering
    \hspace*{-0.8cm}
    \includegraphics[scale = 0.58]{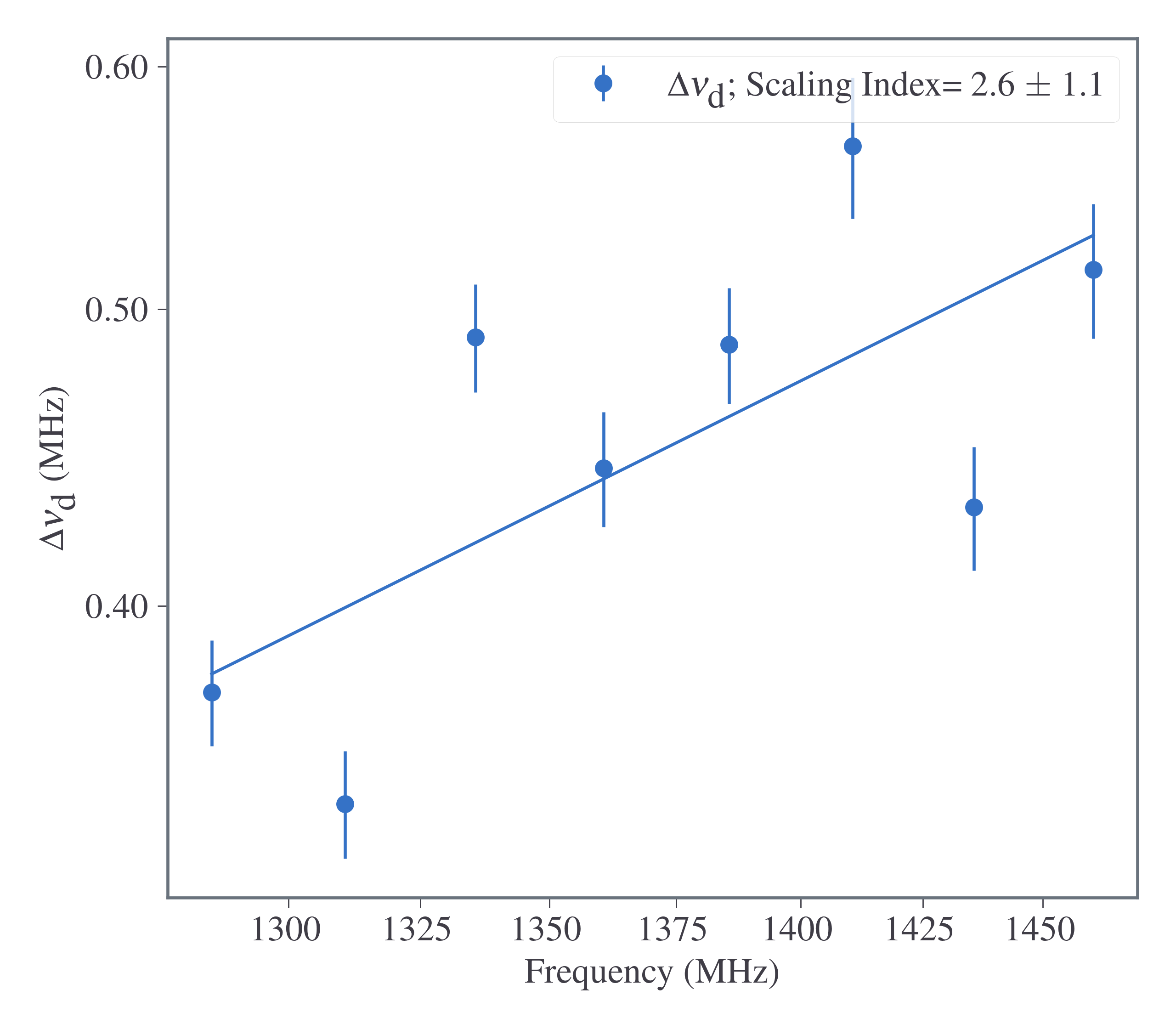}
    \caption{Scintillation bandwidth scaling index fit (blue line) using the weighted average of all scintillation bandwidth measurements in a given subband across all three epochs, with 1$\sigma$ weighted uncertainties shown. Both axes are in log space.}
    \label{avg_bw}
\end{figure}

 Thanks to our ability to fit eight simultaneous measurements across the observing band, our scintillation bandwidth scaling analyses are robust in the context of a PTA-style, i.e., with an emphasis on time resolution, usually at the expense of frequency resolution, observation at L-band, as similar efforts on data with current NANOGrav frequency resolution yield scintillation bandwidths estimations that are 2$-$3 times larger than the true scintillation bandwidth and scaling indices that are consistent with zero within error.  The results are shown in Table \ref{scattering_scaling_results}, with the corresponding fits shown in Figure \ref{ex_scale_fit_delay}. Taking a weighted average of each subband across all three epochs and then performing the fit results in a cross-epoch scaling index of 2.6 $\pm$ 1.1. This weighted average fit is shown in Figure \ref{avg_bw}.

\begin{deluxetable}{CC}[!ht]
\tablewidth{0pt}
\tablecolumns{2}
\tablecaption{Fitted Pulsar Scintillation Bandwidth Scaling Indices \label{scattering_scaling_results}}
\tablehead{\colhead{MJD} & \colhead{\hspace{3.5cm} $x_{\Delta \nu_{\rm d}}$}\hspace{.5cm}}
\startdata
56183 & \hspace{3.5cm}1.3 $\pm$ 1.1\hspace{.5cm} \\ 
56198 & \hspace{3.5cm}4.3 $\pm$ 1.7 \hspace{.5cm}\\ 
56206 & \hspace{3.5cm}3.1 $\pm$ 1.4\hspace{.5cm}\\
\enddata
\tablecomments{Fitted scintillation bandwidth scaling indices to the eight frequency ACFs across the observing band. Errors are 1$\sigma$ fit uncertainties.}
\end{deluxetable}

\par We note that the figures shown in this subsection exhibit a high degree of spread in the data surrounding the fits, and that the resulting scaling indices are often not well constrained. After experimenting with both four and eight-subband fits, we found the measured scaling indices to be relatively insensitive to the number of subbands used, and had mixed results achieving more constrained, higher quality fits when using fewer subbands. We believe underestimated errors may at least partly explain the quality of these fits.

\par Our scaling indices all agree with each other within error and are consistent with those previously reported in the literature for this pulsar, with sources generally quoting indices between 3$-$3.6 within 1$\sigma$ uncertainty both on indices averaged over many epochs as well as for individual epochs \citep{cordes_1937, ramachandran_2006,Levin_Scat,turner_scat,epta1}, although \cite{Levin_Scat} reported indices on individual days as shallow as 1.3 $\pm$ 0.3 and as steep as 4.9 $\pm$ 0.8. 


\par
\begin{figure*}[!ht]
    \captionsetup[subfigure]{labelformat=empty}
    \subfloat[\centering]{\hspace*{-.45cm} {\includegraphics[width=0.36\textwidth]{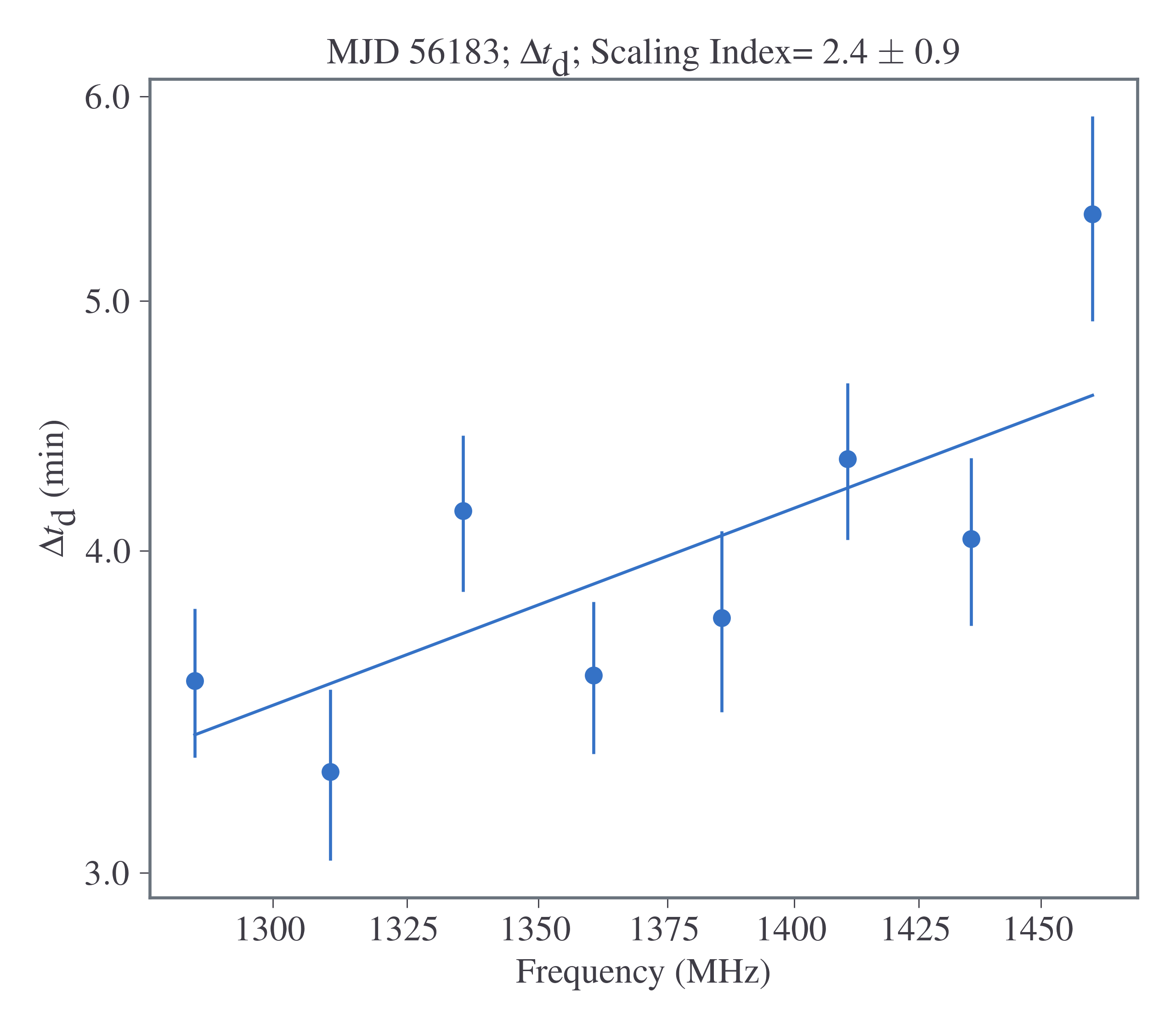} }}%
    \subfloat[\centering]{\hspace*{.0cm} {\includegraphics[width=0.36\textwidth]{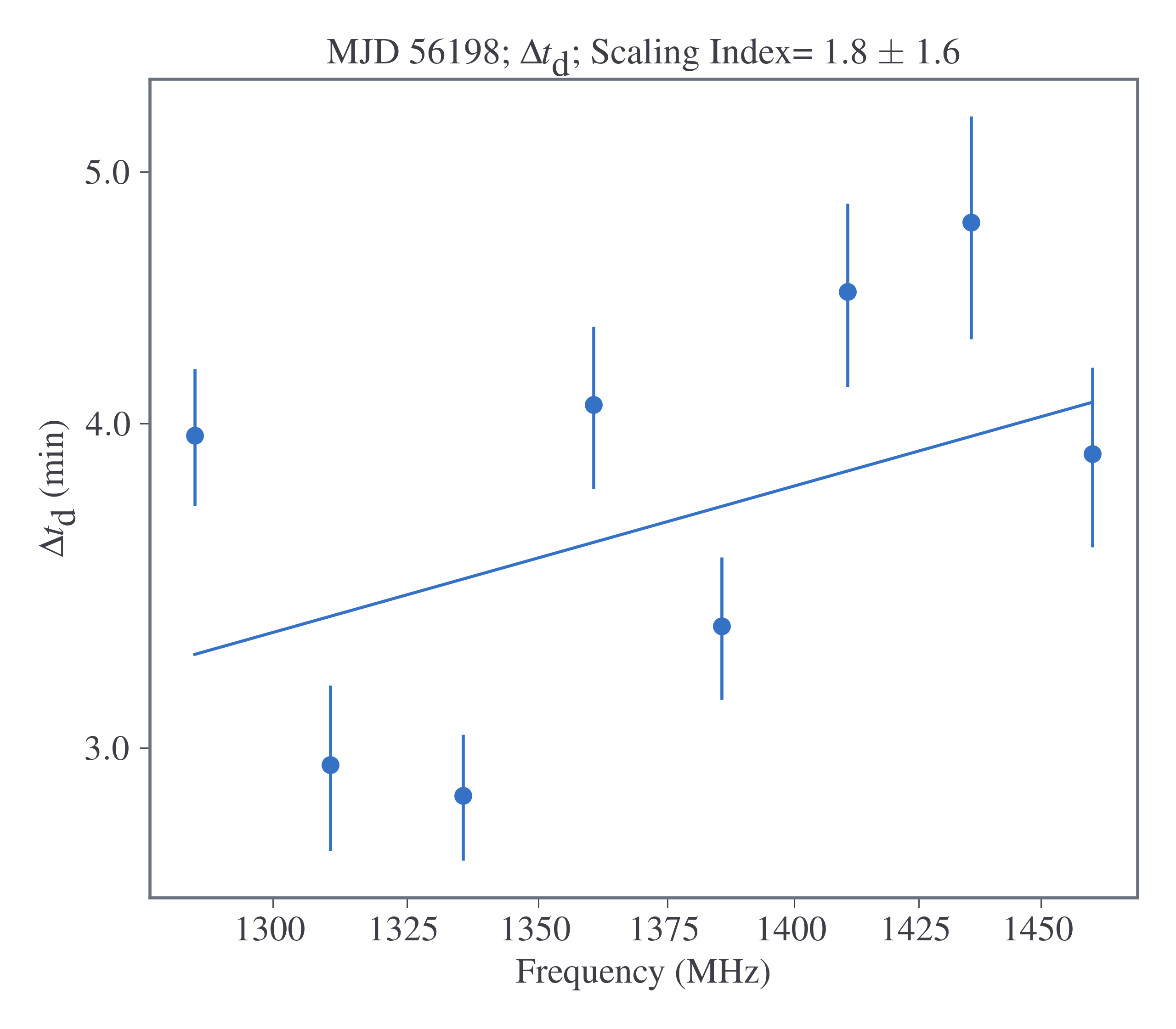} }}%
    \subfloat[\centering]{\hspace*{.1cm}{\includegraphics[width=0.36\textwidth]{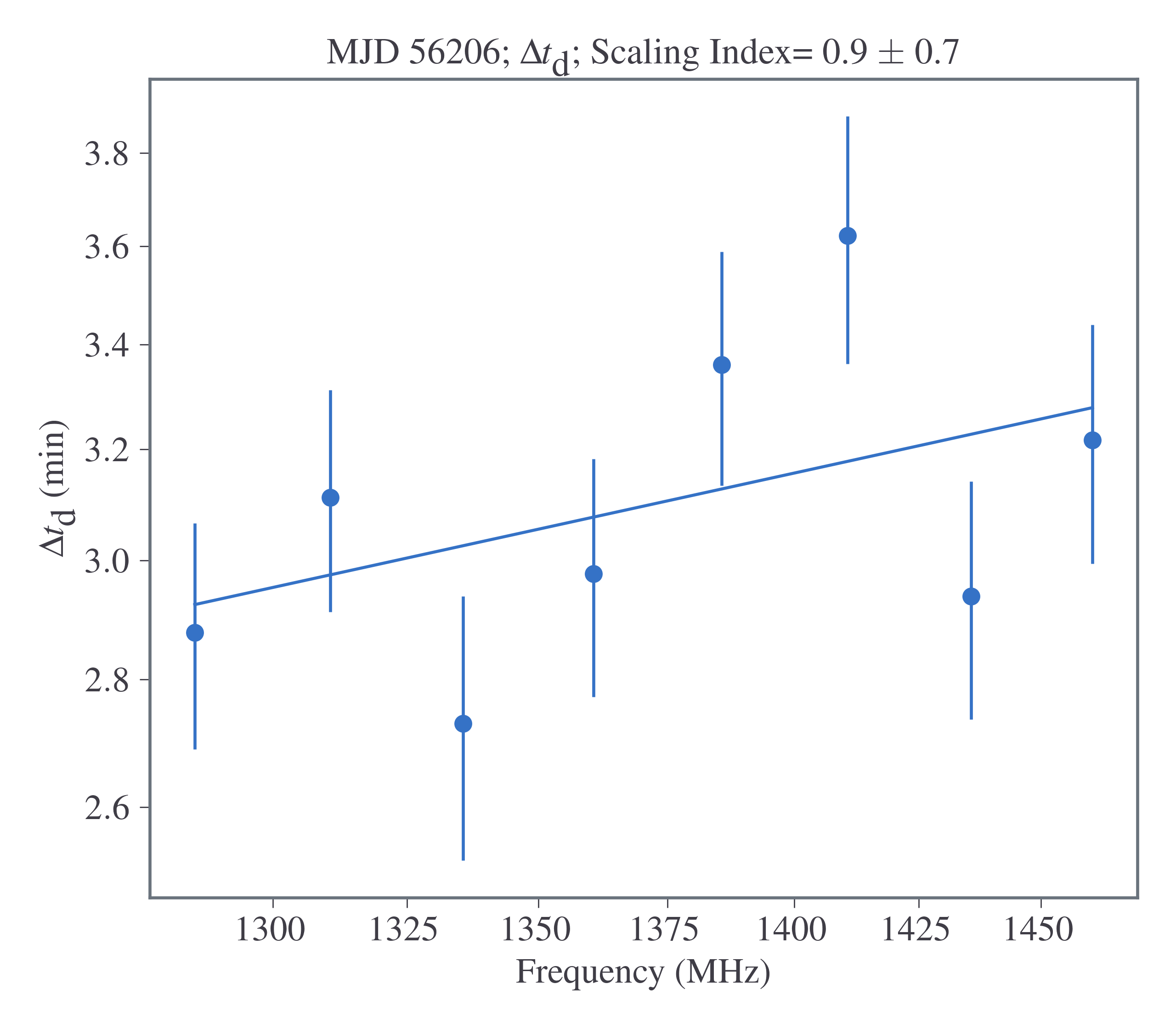} }}%
    \caption{Scintillation timescale scaling indices, from left to right, on MJDs 56183, 56198, and 56206, with scaling index fits indicated by the blue lines. 1$\sigma$ error bars are shown. Both axes are in log space on all plots.}%
    \label{ex_scale_fit_timescale}%
\end{figure*}

\begin{figure}[t]
    \centering
    \hspace*{-.8cm}
    \includegraphics[scale = 0.58]{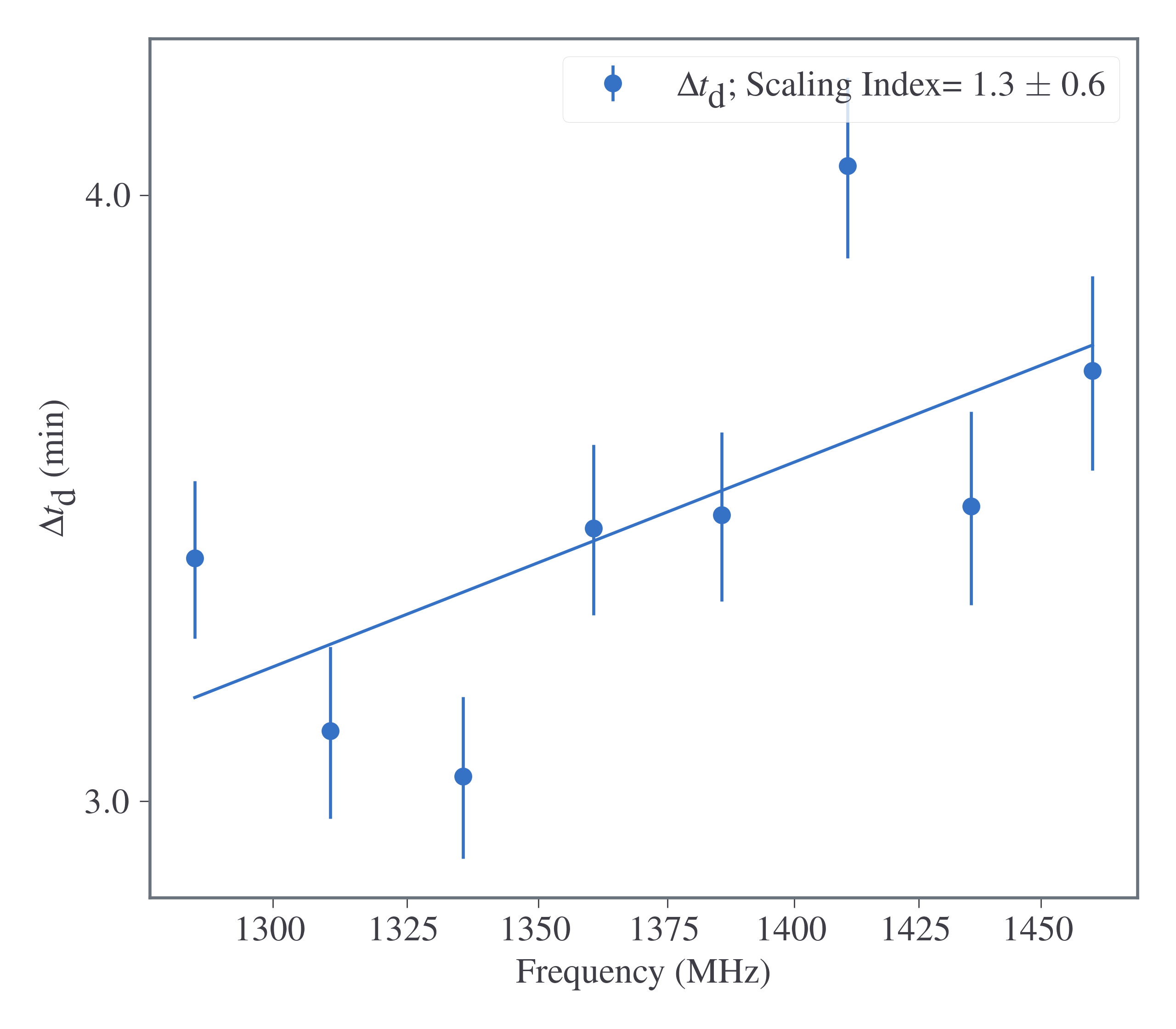}
    \caption{Scintillation timescale scaling index fit (blue line) using the weighted average of all scintillation timescale measurements in a given subband across all three epochs, with 1$\sigma$ weighted uncertainties shown. Both axes are in log space.}
    \label{avg_ts}
\end{figure}

 As a consequence of the longer observing times required to measure scintillation timescales for most pulsars, few studies examine the corresponding scaling behavior. The studies that do exist use measurements taken on separate days \citep{timescale_long}, and simultaneously at multiple frequencies \citep{Bhat_2018, turner_arcs}, although the latter has primarily been done at observing frequencies considerably below 1.4 GHz, where scintillation timescales are much shorter. While a pulsar like PSR B1937+21 has a relatively short scintillation timescale (around 7 minutes at 1.4 GHz per \cite{cordes_1937} and around 3$-$5 minutes in this work), making it possible to track using PTA-length observing campaigns, the large scattering delays make resolving scintles difficult under these setups. Our scintillation timescale scaling indices are shown in Table \ref{timescale_table}, with the corresponding fits in Figure \ref{ex_scale_fit_timescale}. Taking a weighted average of each subband across all three epochs and then performing the fit results in a cross-epoch scaling index of 1.3 $\pm$ 0.7. This weighted average fit is shown in Figure \ref{avg_ts}.

\begin{deluxetable}{CC}[!ht]
\tablecolumns{2}
\tablecaption{Fitted Pulsar Scintillation Timescale Scaling Indices \label{timescale_table}}
\tablehead{\colhead{MJD} & \colhead{\hspace{3.5cm} $x_{\Delta t_{\rm d}}$}\hspace{.5cm}}
\startdata
56183 & \hspace{3.5cm}2.4 $\pm$ 0.9 \hspace{.5cm} \\ 
56198 & \hspace{3.5cm}1.8 $\pm$ 1.6 \hspace{.5cm}\\ 
56206 & \hspace{3.5cm}0.9 $\pm$ 0.7\hspace{.5cm}\\
\enddata
\tablecomments{Fitted scintillation timescale scaling indices to the eight frequency ACFs across the observing band. Errors are 1$\sigma$ fit uncertainties.}
\end{deluxetable}

\subsection{Scintillation Arcs}\label{arc_sec}
\begin{figure*}[!ht]
    \centering
    \captionsetup[subfigure]{labelformat=empty}
    \subfloat[]{ {\hspace{-1cm}\includegraphics[width=0.51\textwidth]{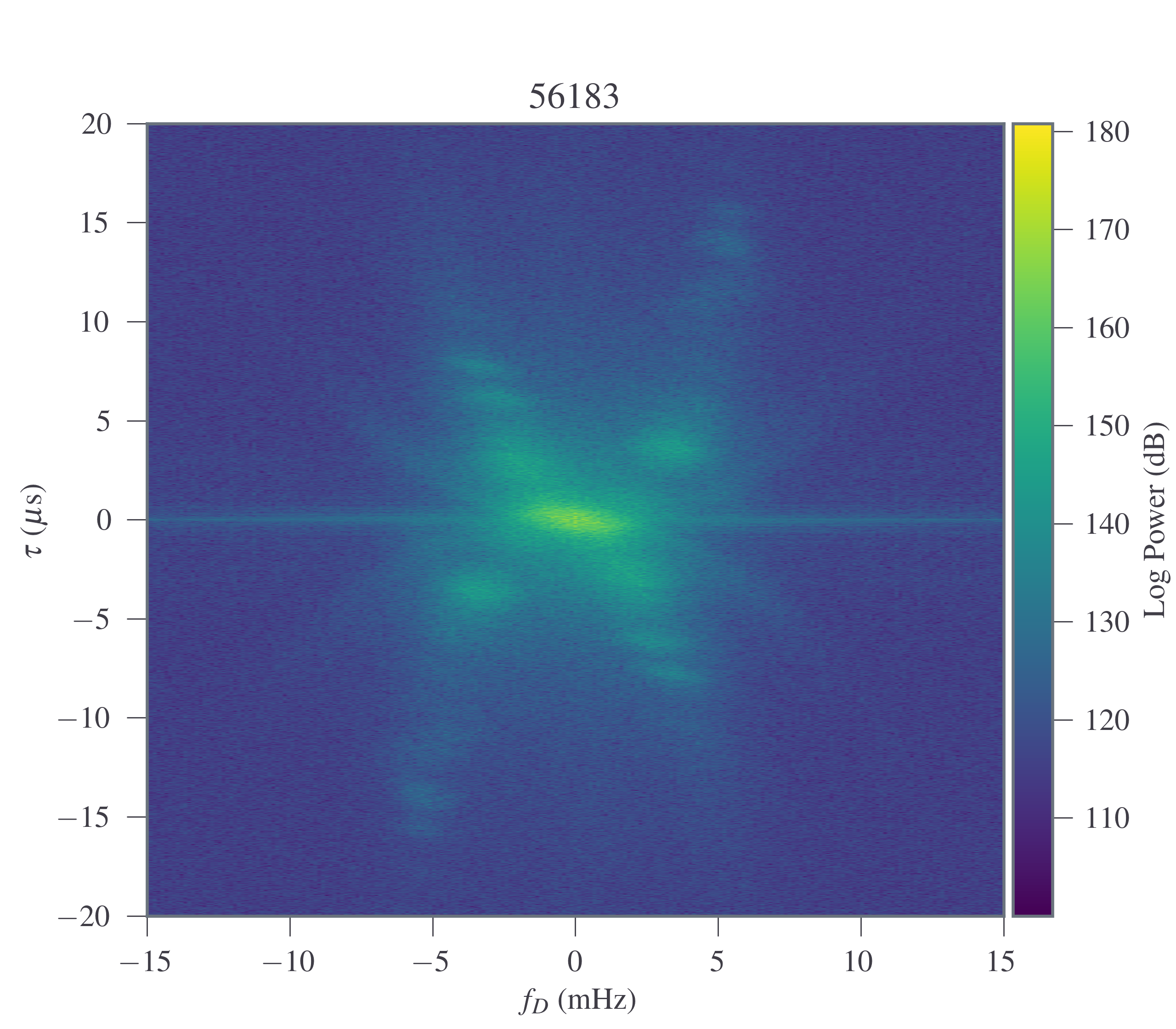} }}\quad
    \subfloat[]{ {\hspace{0cm}\includegraphics[width=0.51\textwidth]{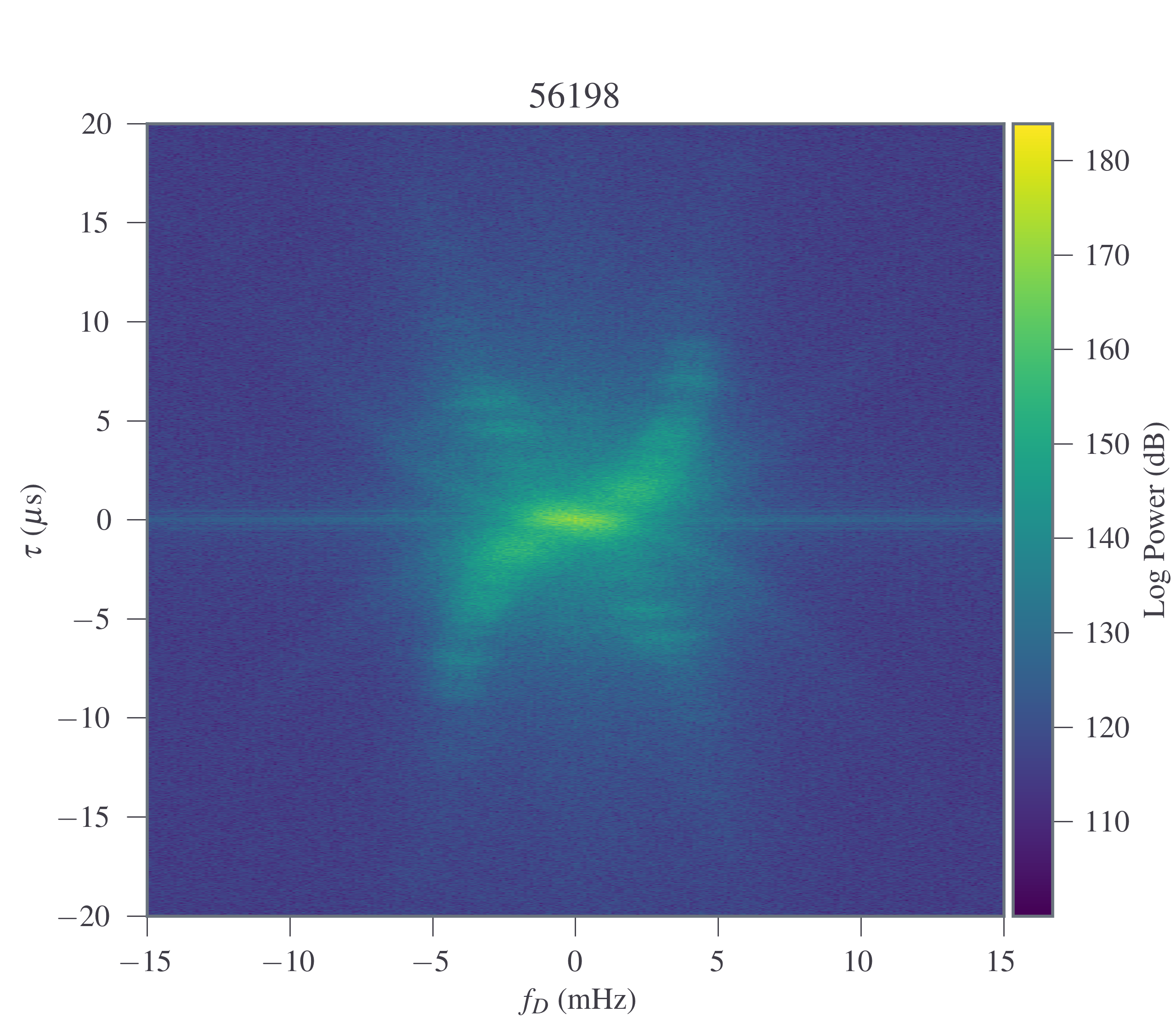} }}\\
    \subfloat[]{{\includegraphics[width=0.51\textwidth]{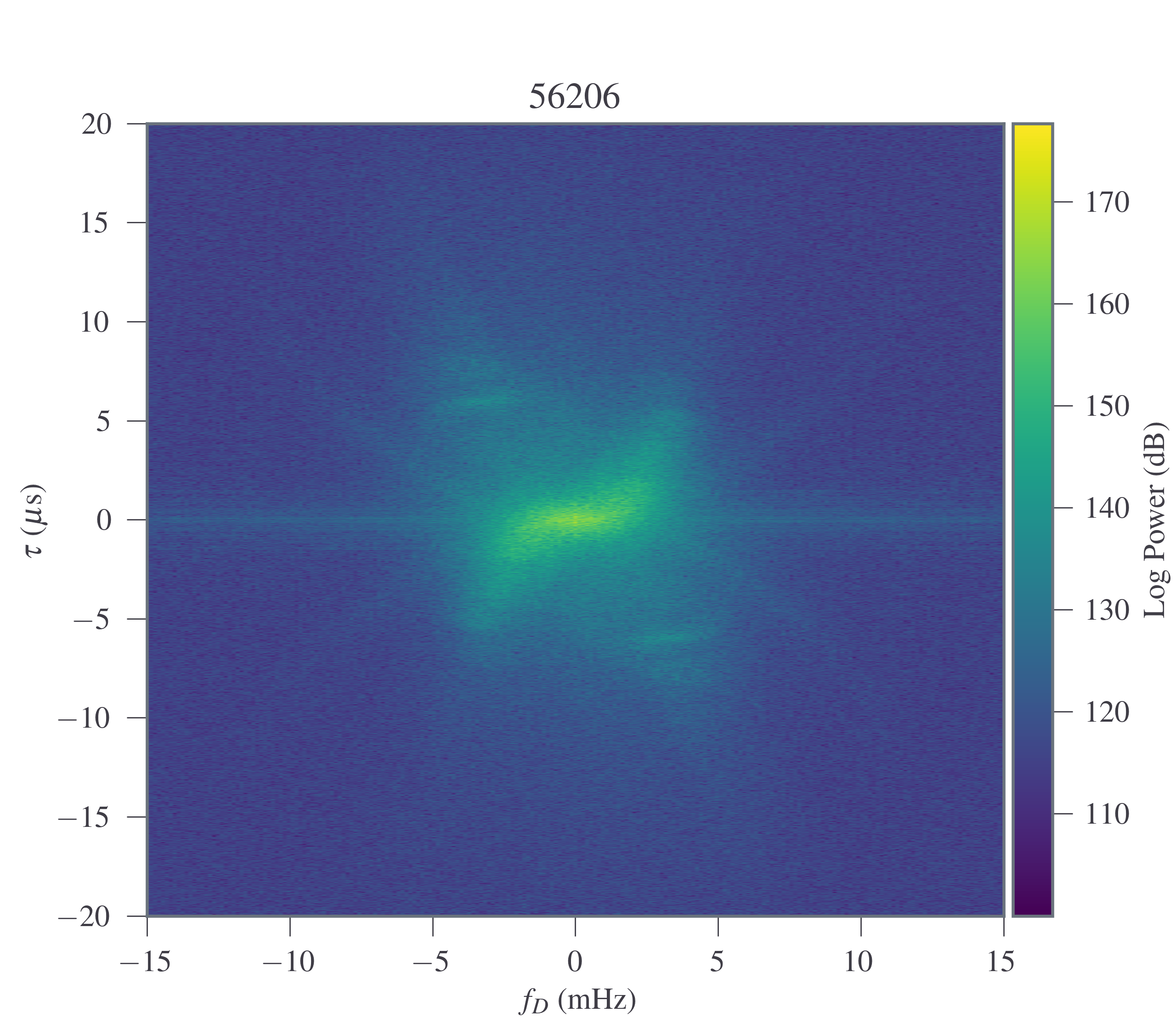} }}%
    \caption{Scintillation arcs on MJDs 56183, 56198, and 56206.}%
    \label{all_arcs}%
\end{figure*}

Scintillation arcs from each of our three epochs can be seen in Figure \ref{all_arcs}. With our high frequency resolution measurements, we can obtain secondary spectra with conjugate frequencies $\tau$ (also called delay) that extend far beyond what is traditionally possible for PTA-style observations, i.e., time and frequency resolutions that are limited by the pulse width and sampling rates. This is a consequence of the Nyquist limit for delay, which is given by $\tau_{\rm Nyq} = N_{\rm chan}/2\rm{BW}$, where $N_{\rm chan}$ is the number of frequency channels and BW is the observing bandwidth. The difference in both the visible range and detail within spectra can be seen quite clearly when comparing secondary spectra created using data processed with a frequency resolution typical of NANOGrav observations and secondary spectra processed using CS, as in Figure \ref{ss_compare}. Of particular significance is that in standard NANOGrav observations we are only sensitive to $\tau$ values less than 0.3 $\mu$s, whereas our CS-processed data extends out to over 80 $\mu$s. However, it is important to note that we have limited the range in our plots to 20$\mu$s, as the range over which arcs are visible in our data only extend out to close to this delay. Comparing these two arcs, we can see that the secondary spectrum processed at current NANOGrav frequency resolution only contains features associated with the power near the origin of the secondary spectrum in the CS-processed data, and so neither of the arms of the arc, much less features within the arms, are visible.

\par
\begin{figure*}[!ht]
    \subfloat[\centering Secondary spectrum on MJD 56183 at 1310 MHz, with the fits to both arms in the scintillation arc shown in orange.]{\hspace*{-.8cm} {\includegraphics[width=0.55\textwidth]{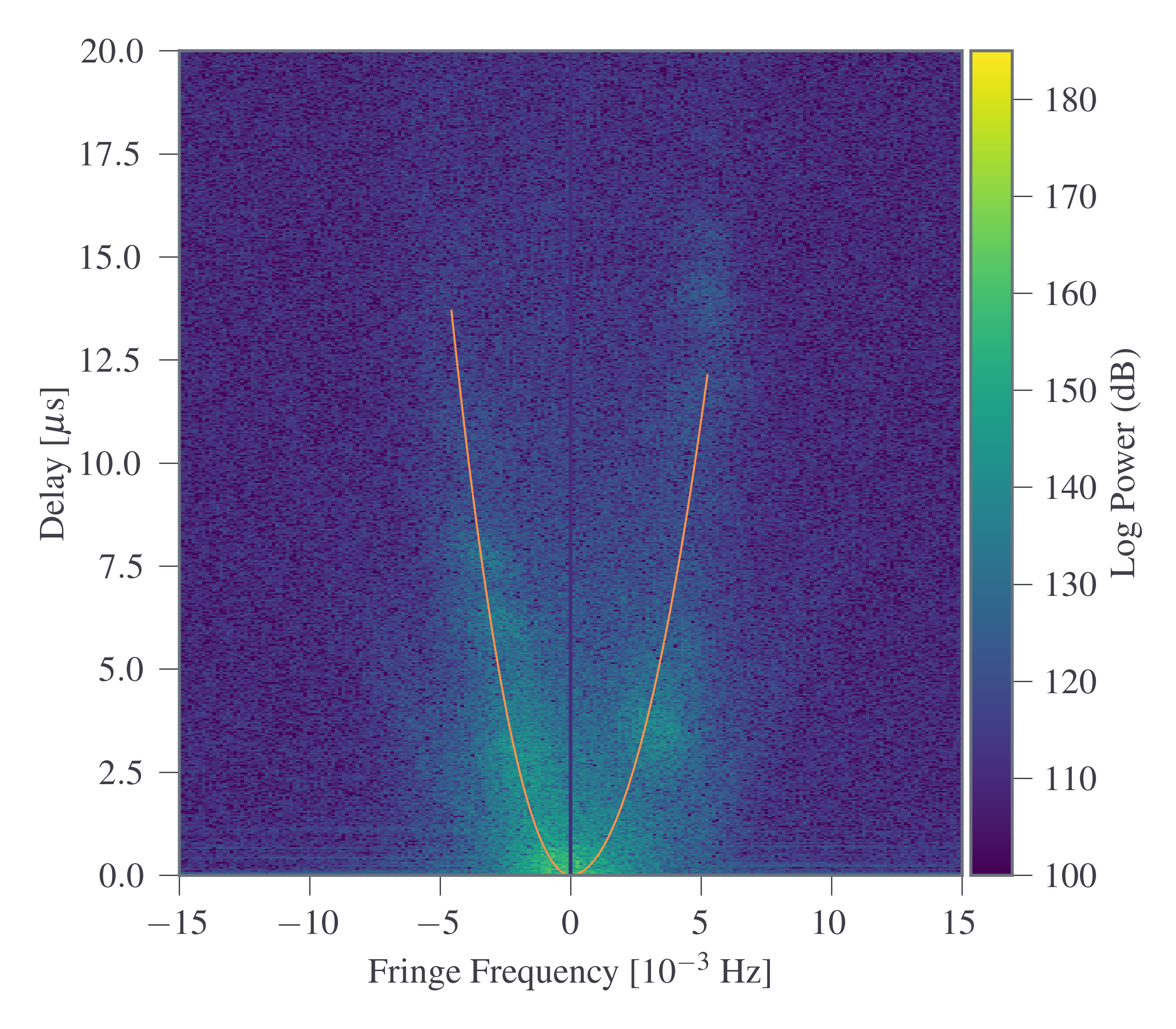} }}%
    \subfloat[\centering Secondary spectrum on MJD 56198 at 1310 MHz, with the fits to both arms in the scintillation arc shown in orange.]{{\includegraphics[width=0.55\textwidth]{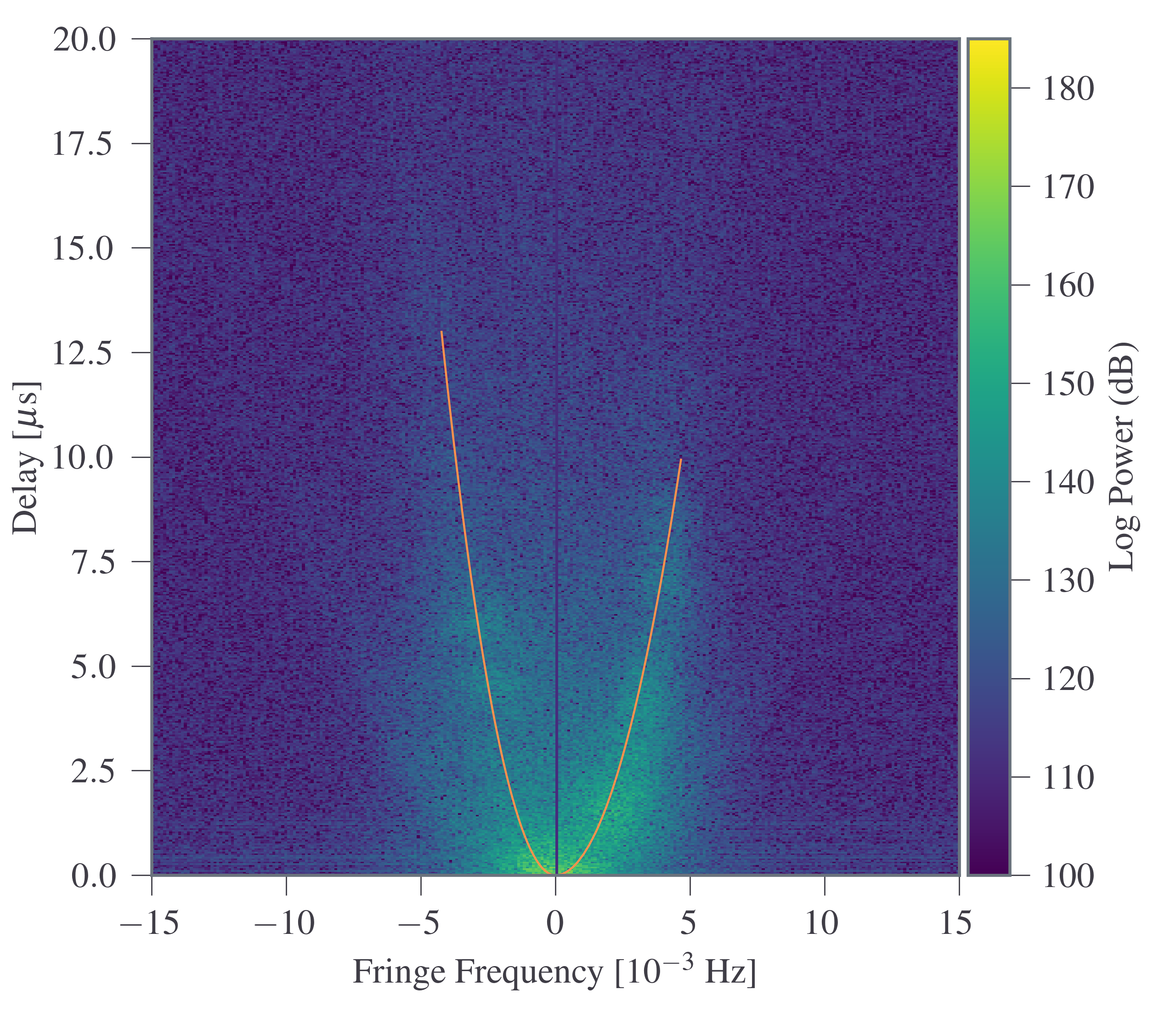} }}%
    \caption{Example scintillation arcs with overlaid fits.}%
    \label{ex_arc_fits}%
\end{figure*}

 Additionally, we have the ability to analyze the intra-epoch evolution of scintillation arcs across a wide range of frequencies, as we were able to resolve a scintillation arc in each of our 25 MHz slices across 200 MHz of bandwidth in all epochs. This represents a substantial improvement in arc sensitivity and indicates the use of cyclic spectroscopy in future studies of MSPs could lead to important developments in our understanding of both inter- and intra-epoch arc evolution, particularly when observed over the large bandwidths used in PTAs. Some example scintillation arcs with overlaid fits can be seen in Figure \ref{ex_arc_fits}.

\subsubsection{Arc Features \& Evolution Over Frequency}
\par Scintillation arcs for this pulsar have been previously shown once in the literature at 427 MHz also utilizing cyclic spectroscopy (MJDs 53791, 53847, and 53873) \citep{1937_cs} and once at L-band using Fourier spectroscopy filterbank methods (MJDs $\sim$56250$-$59000) \citep{main_leap}. Depending on the epoch, the arc power seen in \cite{main_leap} varies between significant asymmetry and roughly symmetric, while the arcs in both \cite{1937_cs} and this work all display noticeable asymmetry. Additionally, at least two arcs can be seen in both their data and ours. An important distinction between the arcs in \cite{1937_cs} and those in both our work and \cite{main_leap} is the significant difference in the scale of our fringe frequency axes. Using a by-eye comparison and the relation between fringe frequency, observing wavelength, and angle on the sky, $\theta \propto f_D\lambda$ \citep{hsa+05}, we have shown the arcs in our data probe an angular scale 6$-$10 times smaller than the arcs in \cite{1937_cs}. Additionally, due to the significantly larger delay axes in \cite{1937_cs}, all structures within the arcs we observe are fully contained within the central feature of their arcs.
\par Arclets, resolved as patchy features in our scintillation arcs, which have been attributed to a variety of astrophysical phenomena, including AU-scale inhomogeneities within the scattering screen \citep{hsa+05} or a double lensing effect from material behind the scattering screen relative to Earth \citep{mult_screen_1508, Zhu_2023}, can be seen in the secondary spectra in our data, and indeed can be seen in all of the observed scintillation arcs in both arms on MJDs 56183 and 56198 and in the left arm on MJD 56206. These were also seen in the arcs shown in \cite{1937_cs}, although only visible in the right arm, which is also the brighter arc in those observations, but not in \cite{main_leap}. They mention their arcs are largely featureless, although they do report some discrete structures in a single epoch. This makes our observations the first reported detections of arclets in this pulsar at L-band. Additionally, while \cite{main_leap} had similar frequency resolution and observed in the same frequency range as our data, their lack of arclets is likely a consequence of lower S/N and shorter observing length; their observations of PSR B1937+21 were only around 50 minutes, whereas ours were around 150. As in \cite{Hill_2003}, we expect arcs at lower frequencies to be wider and have more diffuse features. While difficult to discern throughout the evolution in the arcs in our data, when comparing with the secondary spectrum seen in the 427 MHz observations in \cite{1937_cs} the arcs displayed in that spectrum are clearly wider and more diffuse compared to the arcs in our data when examined by eye (around 6-8 mHz in width for their data compared to around 2-3 mHz in width in ours), agreeing with this expectation. 
\begin{figure*}[htbp]
\centering
\includegraphics[width=0.36\textwidth]{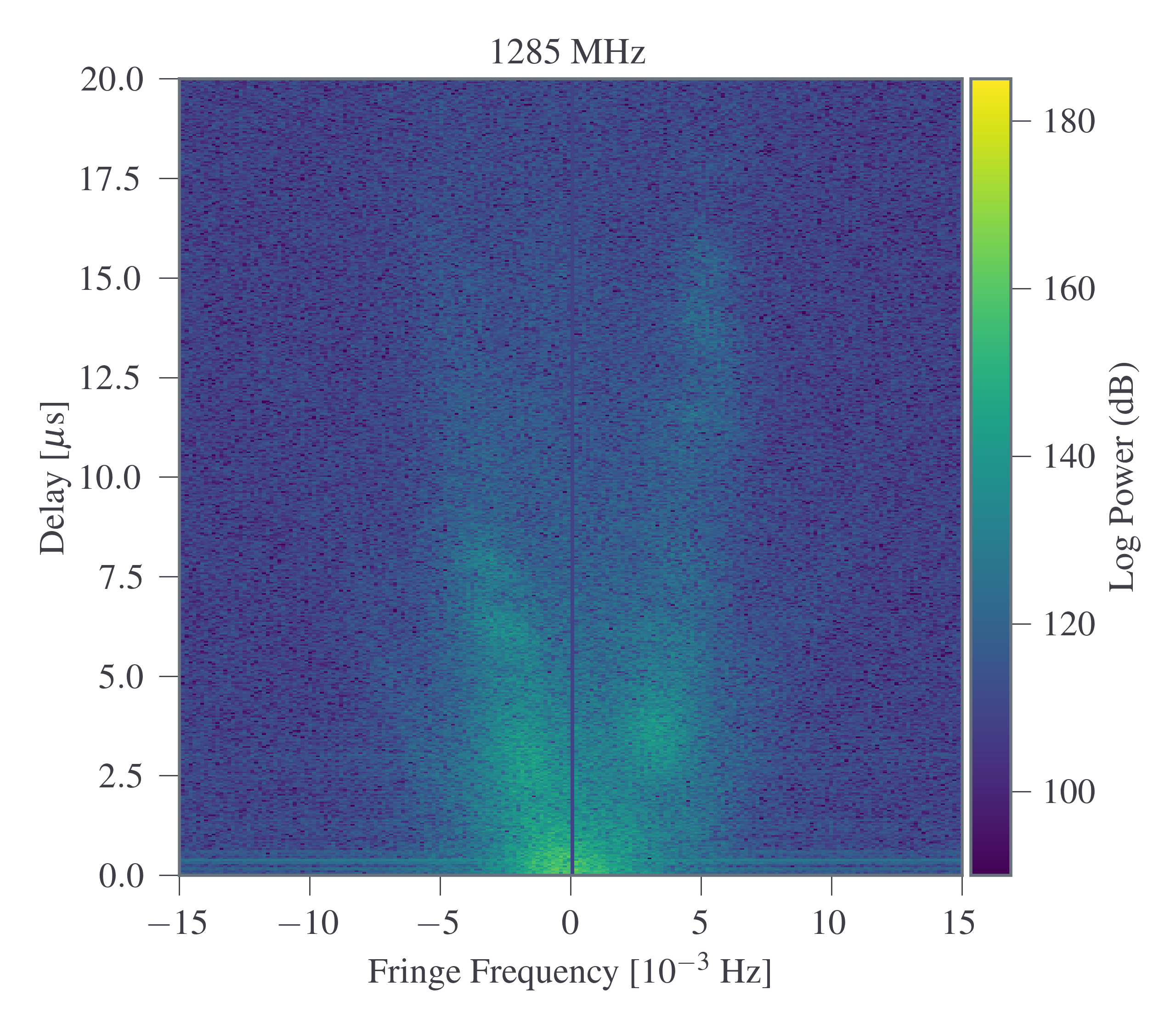}
\hfil
\includegraphics[width=0.36\textwidth]{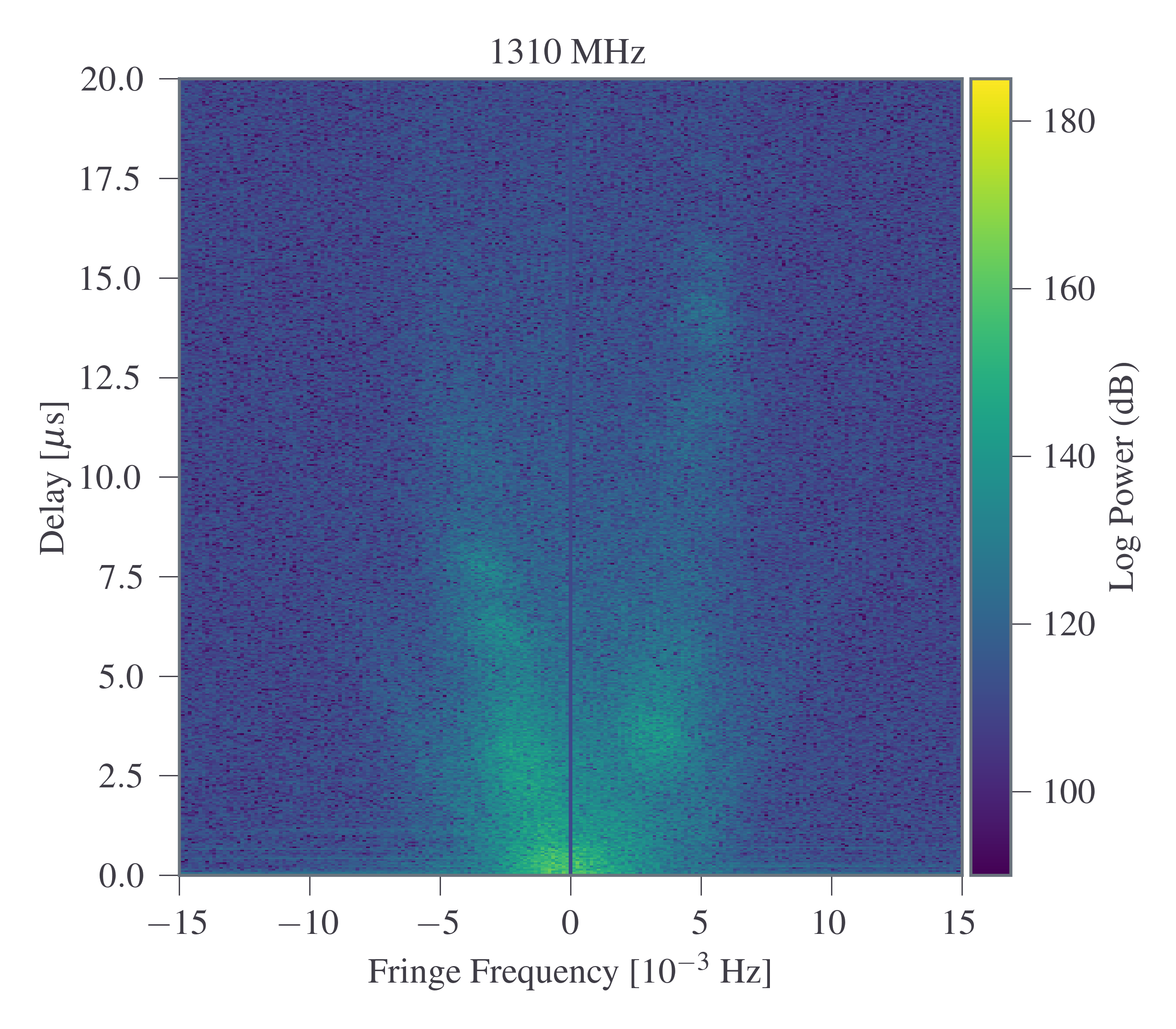}

\medskip
\includegraphics[width=0.36\textwidth]{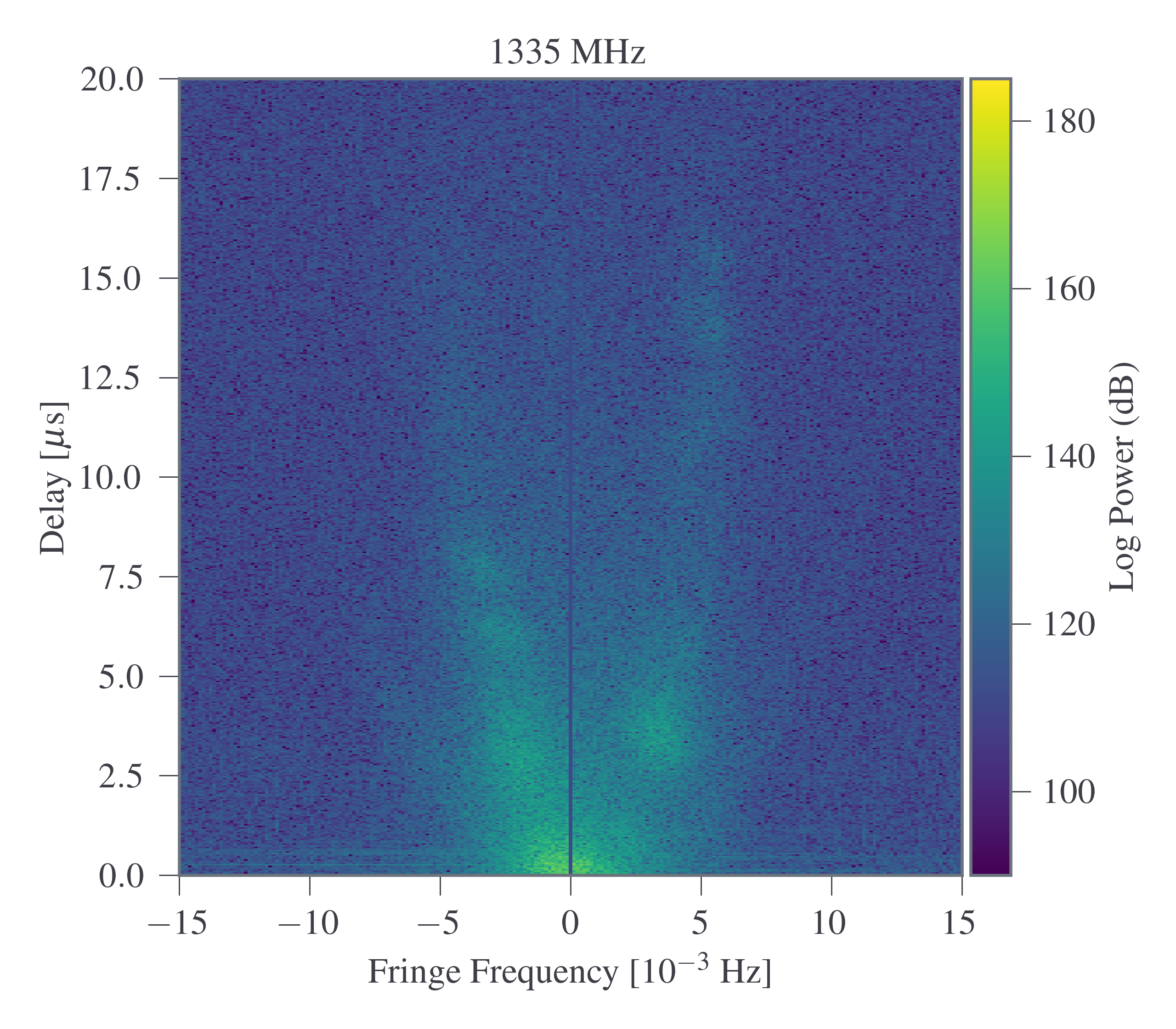}
\hfil
\includegraphics[width=0.36\textwidth]{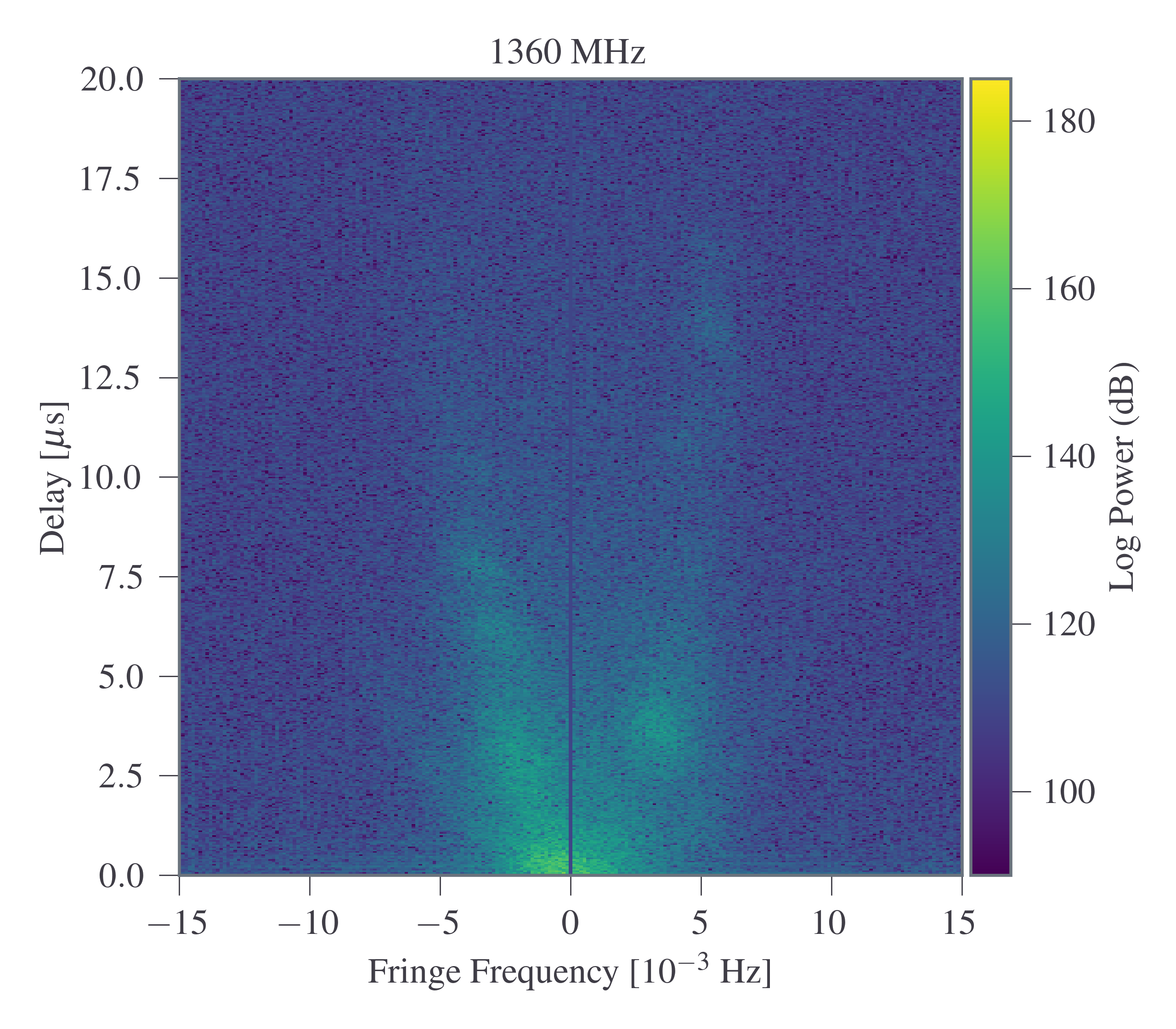}

\medskip
\includegraphics[width=0.36\textwidth]{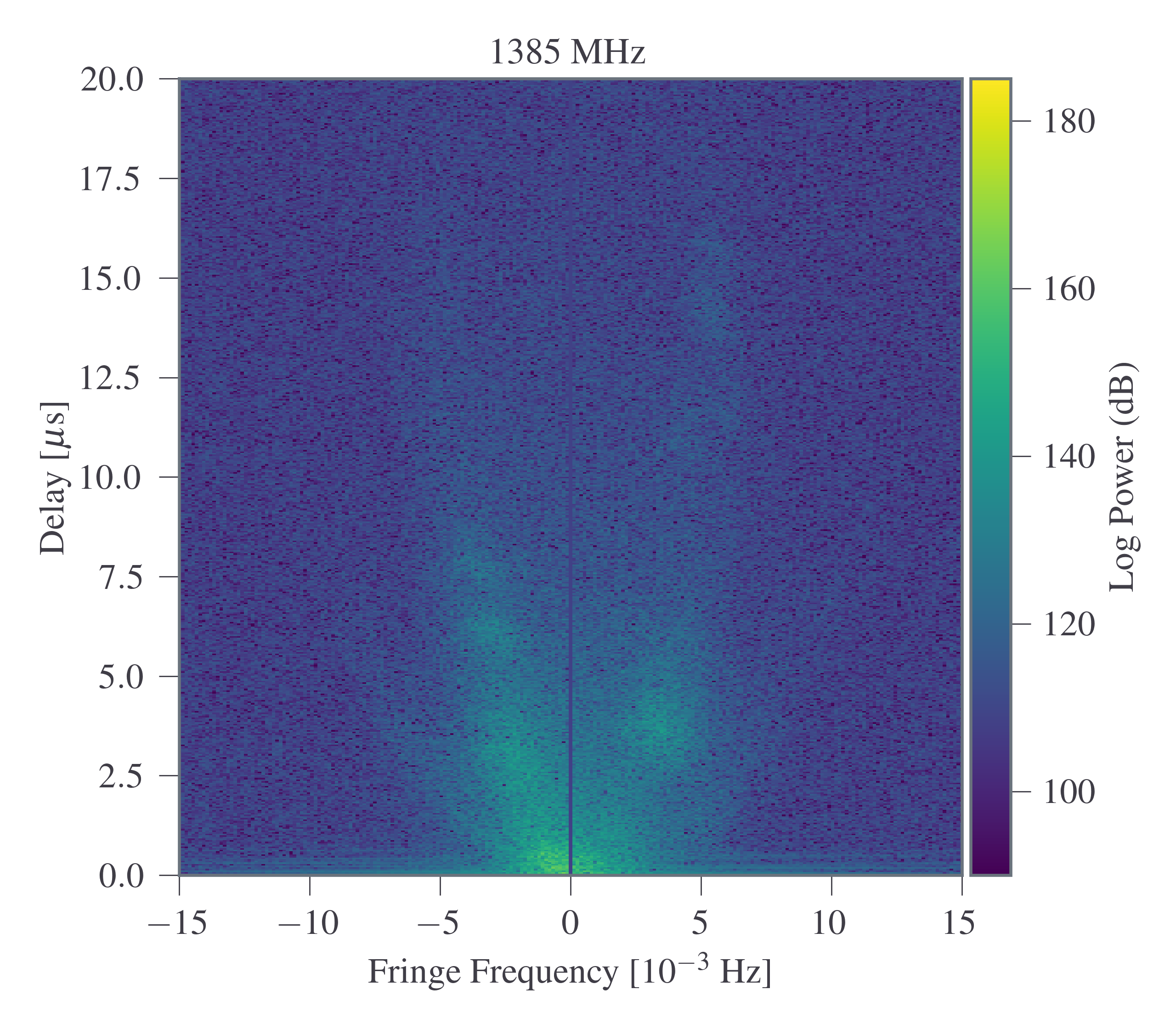}
\hfil
\includegraphics[width=0.36\textwidth]{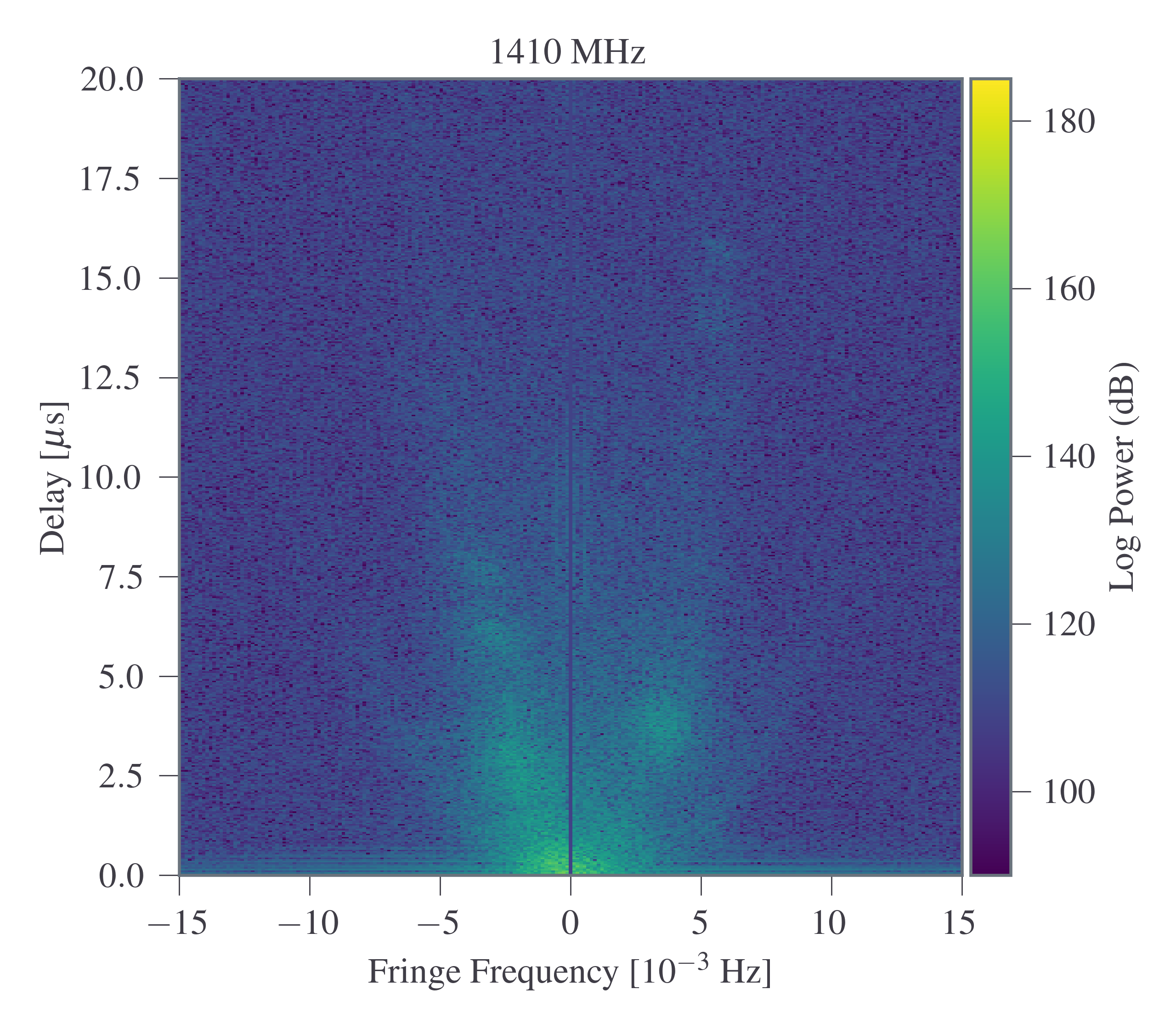}

\medskip
\includegraphics[width=0.36\textwidth]{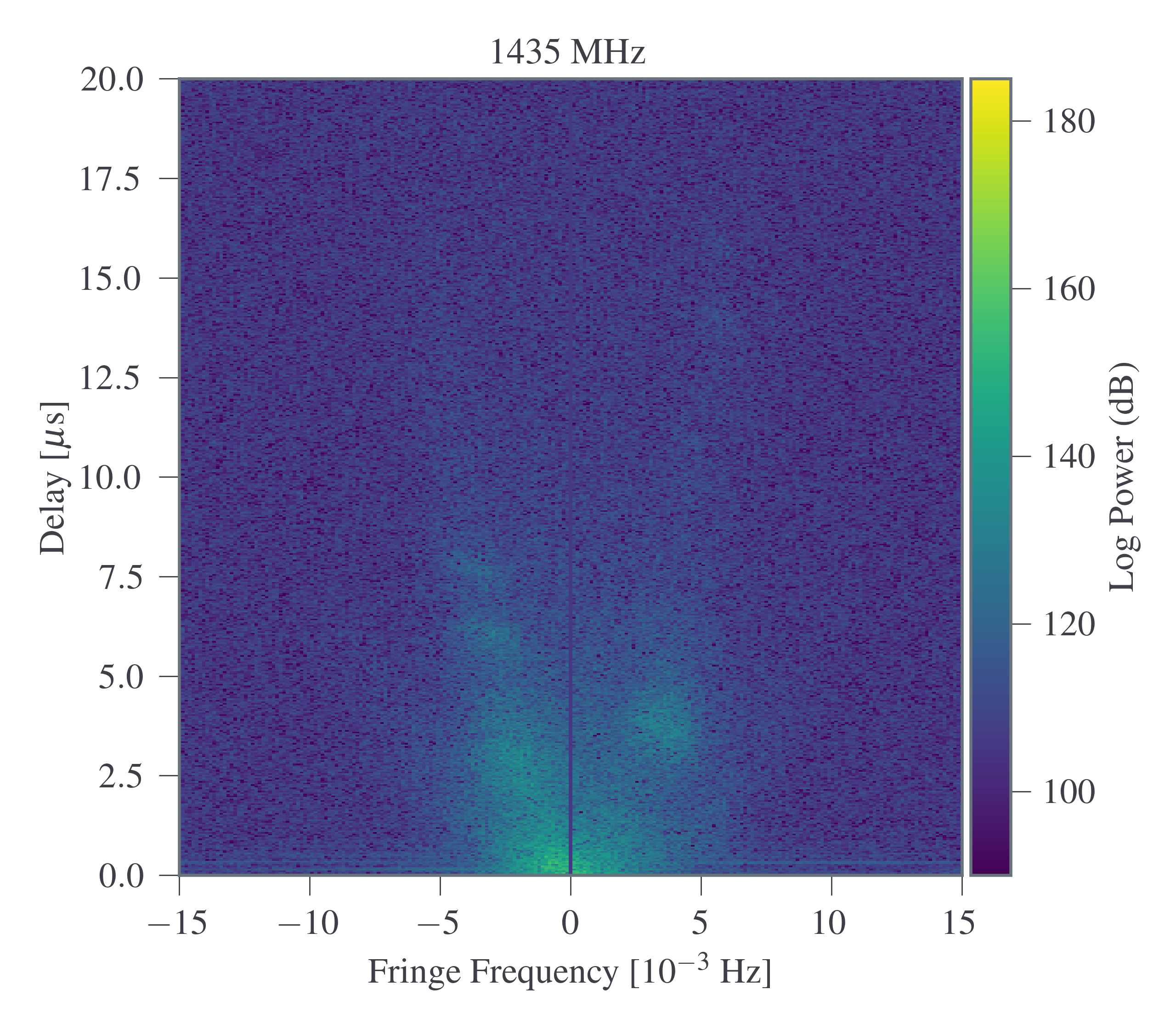}
\hfil
\includegraphics[width=0.36\textwidth]{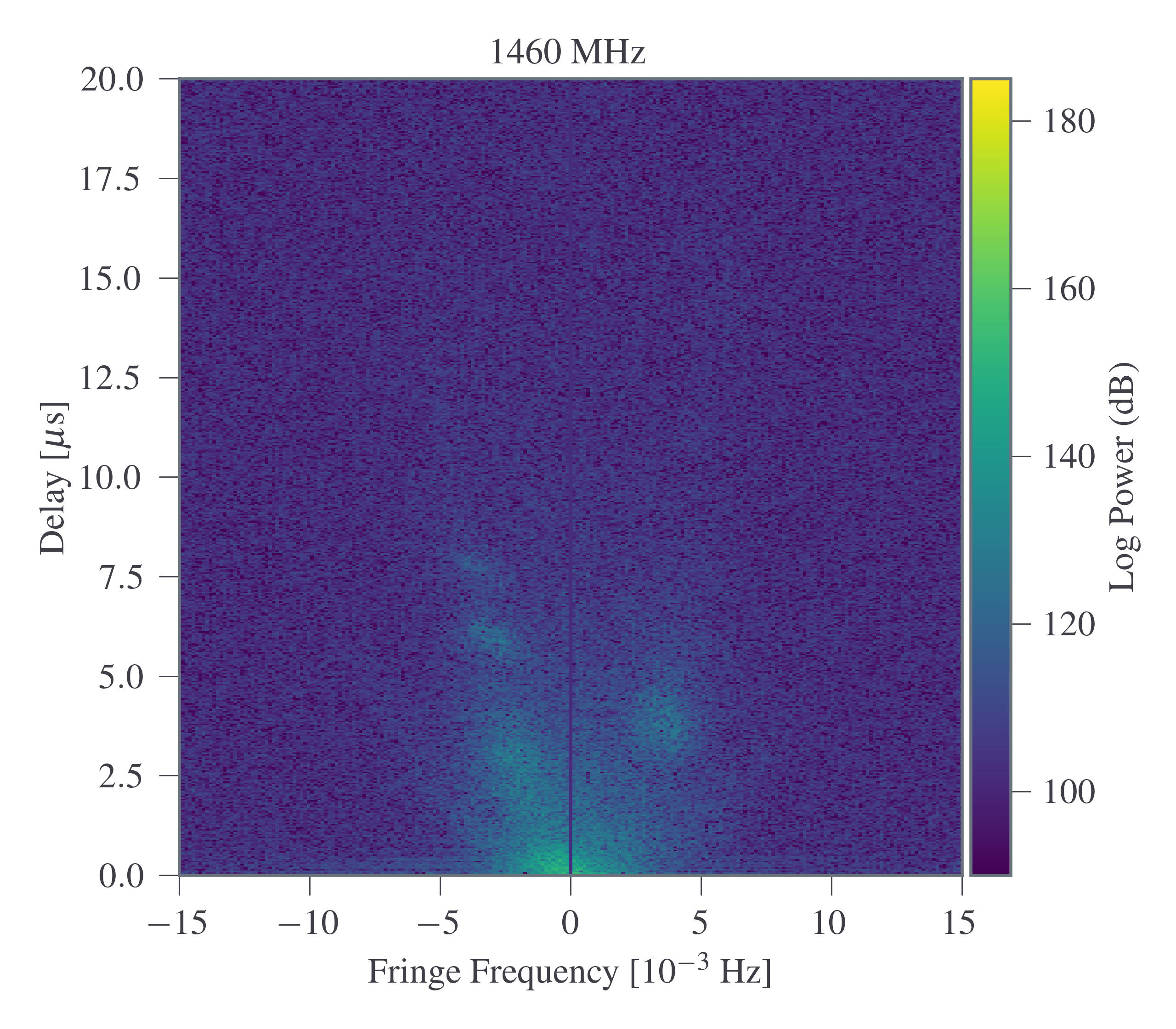}
\caption{Scintillation arc evolution on MJD 56183 in increasing 25 MHz intervals across and down beginning at 1285 MHz (top left) and ending at 1460 MHz (bottom right). Features in the arcs dim as we go to higher frequencies, with only the power near the base of the parabolas and centers of the arclets remaining visible at the highest frequency.}
\label{arc_evo}
\end{figure*}
\par At all three epochs, we find that the arcs dim with increasing observing frequency. The positions of the arclets remain constant with frequency and dim from the edges to the center of the arcs with increasing frequency. Regardless of observing frequency, we expect the features closest to the centers of arcs to remain the brightest, as they correspond to constructive interference in the larger scintle structure in the dynamic spectrum. The first structures that disappear are the dim portions of the arc higher up the delay axis, while the only remaining structures at the highest frequencies are the apexes of the arclets, which were in fact the brightest structures in the arc aside from the concentrated power at the base of the fringe  frequency axis, which also remains visible. This effect may be explained by the frequency dependence of the plasma refractive index. For a given feature offset from the line of sight by a fixed distance, there is some minimal refractive index required to deflect the signal back to Earth. As the refractive index drops as $\nu^{-2}$, there must then be some maximum frequency for which the image is observed. As the images near the base of the parabola are closer to the line of sight, the minimum refractive index is lower and so the images will still be visible at higher frequencies. It is also well established that pulsar flux decreases with frequency over the ranges we observe \citep{alam2020nanograv}, and we can quantitatively see this in the dynamic spectra in our data, as the mean and median flux density of the 25 MHz frequency slices of these spectra clearly get progressively smaller at higher frequencies. As a result, it may be to a certain extent that we simply are unable to see some dimmer features in the higher frequency arcs as a consequence. Additionally, this pulsar is known for having a particularly steep spectral index \citep{1937_discovery}, making this effect much easier to see. That being said, these observations only span 200 MHz, and so changes in flux density over frequency likely only partially explain what we are seeing in our data. An example of this evolution can be seen in Figure \ref{arc_evo}. Note that, although all plots in this figure display the same dynamic range for ease of visual comparison, these features are still present when we allow their dynamic ranges to vary.

\subsubsection{Curvature Measurements \& Scaling}
\par
\begin{figure}[!ht]
    \centering
    \includegraphics[scale = 0.58]{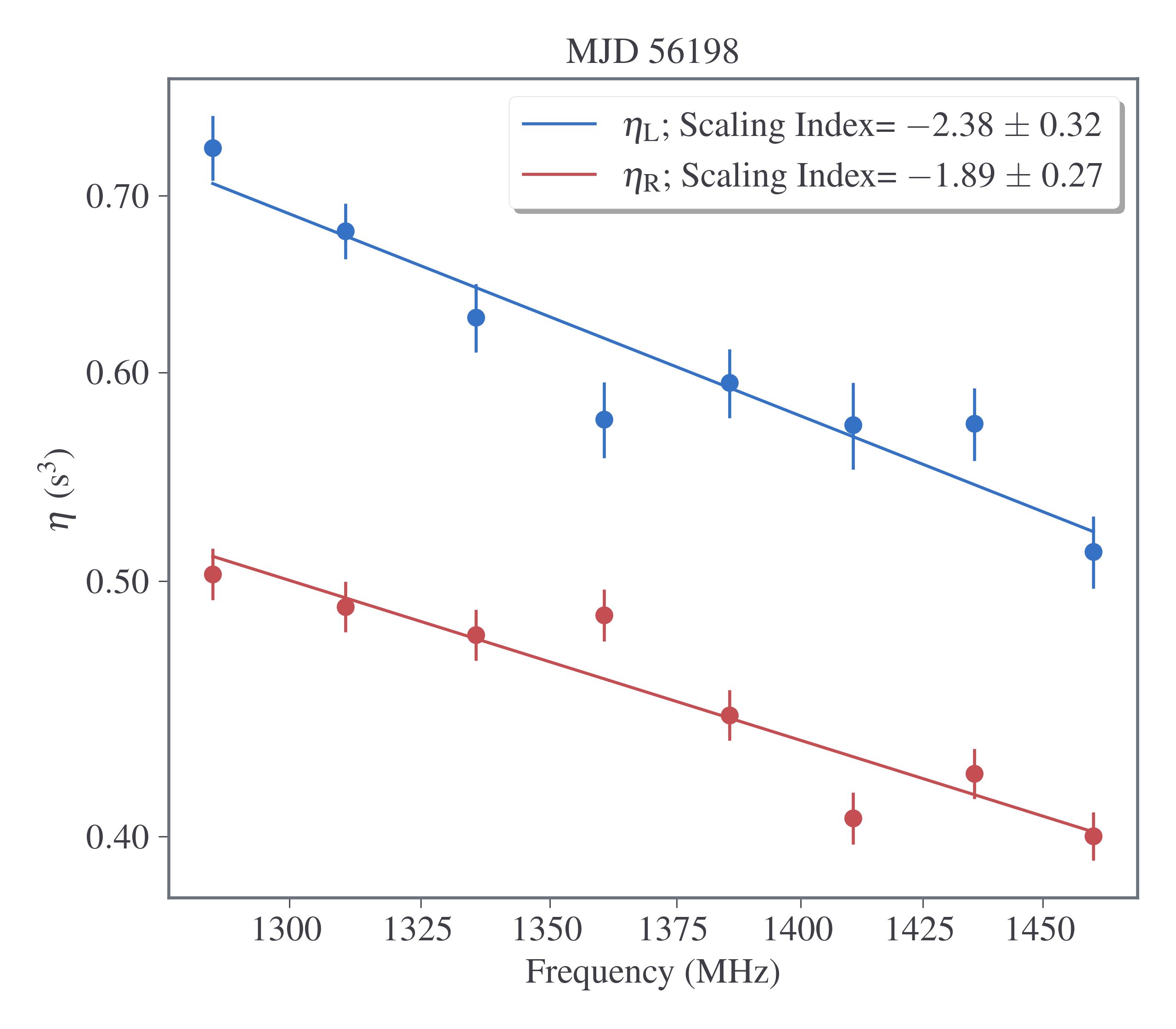}
    \caption{Scintillation arc curvature scaling indices in both arms on MJD 56198. The fits agree well with the frequency dependence of angular deflections as detailed in \cite{Hill_2003}. The offset between the two fits may partially be attributable to difficulties in fitting curvatures to patchy arcs.}
    \label{ex_scale_fit_arcs}
\end{figure}

 The results of our scintillation arc scaling analysis are shown in Table \ref{scaling_results}, with example fits from MJD 56198 shown in Figure \ref{ex_scale_fit_arcs}. We can see that only some of our scaling indices agree with  the $\eta \propto \nu^{-2}$ theory laid out in \cite{Hill_2003}, indicating the curvature of these arcs may not perfectly scale with frequency following the same power law as the angular deflection of the scattered rays.

\begin{deluxetable}{CCC}[!ht]
\tablewidth{0pt}
\tablecolumns{3}
\tablecaption{Fitted Pulsar Scintillation Arc Curvature Scaling Indices \label{scaling_results}}
\tablehead{ \colhead{MJD} & \colhead{$x_{\eta,L}$} & \colhead{$x_{\eta,R}$} }
\startdata
56183 & \text{$--$}2.05 $\pm$ 0.17 & \text{$--$}1.49 $\pm$ 0.33\\
56198 & \text{$--$}2.38 $\pm$ 0.32 & \text{$--$}1.89 $\pm$ 0.27\\
56206 & \text{$--$}1.60 $\pm$ 0.38 & \text{$--$}1.29 $\pm$ 0.28\\
\enddata
\tablecomments{Fitted arc curvature scaling indices for both left and right arcs. Different power laws and curvatures of the left and right arms on a given day may partially be a consequence of the difficulty fitting curvatures to discontinuous arcs comprised mostly of patchy features.}
\end{deluxetable}

\par While our scaling indices are somewhat consistent with $-2$, two epochs display a noticeable asymmetry in the measured index between the left and right arms. Despite the high resolution of these observations, this patchiness may make it difficult to properly measure curvature in all subbands, particularly if the apexes of some arclets are offset from the main parabolic fit, as in \cite{Brisken_2010}. As a result, measured curvatures may slightly deviate from the true curvature of the arcs, which may at least in part contribute to the power law asymmetries we have measured. Consistent offsets in the curvatures of the left and right arms are also visible across all frequencies and all epochs, with curvatures in the right arms always being smaller than those on the left, even when there is more visible structure in the right arm than the left. Interestingly, over the course of these observations, the arc curvatures in right arm transitions from consistently smaller than the left arm to almost identical. An example of these offsets can also be seen in Figure \ref{ex_scale_fit_arcs}. 

Once curvature measurements are obtained, they can be used in conjunction with $\textbf{V}_{\textrm{eff}, \perp}$ to estimate screen distances and potentially associate scattering screens with known structures along the LoS to a pulsar. In cases where the pulsar velocity dominates, we can ignore the velocities of Earth and the ISM when estimating these distances. However, the low velocity of PSR B1937+21 (6 km s$^{-1}$, per \cite{ding_vlbi}) requires us to take these additional velocities into account. Unfortunately, as a consequence, degeneracies in our estimated screen distance solutions (either around 0.1 or 2.9 kpc from Earth, assuming a screen orientation of zero degrees and an ISM velocity of 10 km s$^{-1}$) necessitate high cadence, long-term arc measurements to properly constrain these distances \citep{Reardon_2020}. These long term observations can properly account for the orbital motion of Earth, which causes periodic variations in the observed curvature, and can help break the degeneracy between screen distance and screen orientation angle. However, it is unlikely that there would be any obvious structures, such as an HII region or molecular cloud, along the LOS to this pulsar which we could associate with a screen in the first place. The LOS is mostly devoid of significant structure and predominantly consists of the inter-arm region between the Orion-Cygnus and Carina-Sagittarius spiral arms of the Milky Way, although a small fraction of the material along the LOS can be attributed to the Orion-Cygnus arm \citep{NE2001}. This likely makes interstellar plasma filaments the most plausible candidate for this pulsar's screen(s) \citep{stock}. In general, as more scintillation arc distances are constrained, there is an increasing need to identify such diffuse structures along LoSs with scattering \citep{ocker}.

\section{Conclusions \& Future Work}
\label{concl}
We have used cyclic spectroscopy to improve the frequency resolution of L-band baseband observations of PSR B1937+21 compared to what is feasible in PTA observing setups that rely on limits of non-cyclostationary time and frequency sampling, yielding highly detailed dynamic and secondary spectra and allowing for scintillation parameter estimations within 25 MHz subbands. These results have yielded scintillation bandwidth scaling indices that match previous observations, as well as the first detection of arclets in this pulsar at L-band. We also found arc curvature scaling indices for different arms within individual observations to exhibit noticeable asymmetries. An examination of arc feature evolution over frequency shows a dimming of prominent features as we go to higher frequencies, at least in part likely due to the frequency dependence of the plasma refractive index, as well as this pulsar's steep spectral index. 
\par The thorough analysis in this work made possible through cyclic spectroscopy demonstrates the potential for larger-scale efforts using this technique and adds to a growing case toward its eventual adoption by PTAs  due to its ability to calculate high-resolution dynamic and secondary spectra. Traditional polyphase filterbanks have limited the combined time and frequency resolution of scintillation studies with pulsars, particularly in studies of scintillation arcs because the secondary spectrum is Nyquist limited. This in turn may make more pulsars accessible for scintillation studies at lower observing frequencies. Additionally, the ability to use cyclic spectroscopy to recover a pulsar's PBF through deconvolution \citep{1937_cs} or other methods \citep{turner_cyc} adds another element to these studies, especially given that the PBF may present a more accurate picture of the scattering present in a given observation than the traditional ACF fitting approach \citep{turner_cyc}.
\par Efforts are currently underway at the Green Bank Observatory to implement a real-time or near real-time cyclic spectroscopy backend, with development nearing completion and testing on real data to begin shortly. This will provide exciting opportunities for new science to all astronomers studying periodic or pulsed emission using this facility, as well as aid in the timing efforts of the NANOGrav PTA collaboration, for whom the 100 meter telescope at the observatory plays a significant role in the ongoing study of low-frequency gravitational waves.
\section{Acknowledgements}
\par We gratefully acknowledge support of this effort from the NSF Physics Frontiers Center grants 1430284 and 2020265 to NANOGrav. All of the data processing in this work utilized the resources of the Link computing cluster at West Virginia University. We thank the NANOGrav Collaboration for helpful discussions and suggestions, particularly the Noise Budget Working Group. We thank the AO staff for their assistance during the observations reported in this work, in particular P. Perillat, J. S. Deneva, H. Hernandez, and the telescope operators. TD, MAM, and DRS were partially supported through the National Science Foundation (NSF) PIRE program award number 0968296. TD acknowledges NSF AAG award number 2009468, sabbatical and summer leave funding from Hillsdale College, and the Hillsdale College LAUREATES program. We also greatly thank Ue-Li Pen, Daniel Baker, Dylan Jow, and Ashley Stock for valuable discussion.

\par \textit{Software}: \textsc{dspsr} \citep{dspsr}, \textsc{astropy} \citep{astropy}, \textsc{pypulse} \citep{pypulse}, \textsc{scipy} \citep{scipy}, \textsc{numpy} \citep{numpy}, and \textsc{matplotlib} \citep{matplotlib}.
\bibliography{turner_1937_cs.bib}{}
\bibliographystyle{aasjournal}
\end{document}

%% file: authors.tex
\author[0000-0002-2451-7288]{Jacob E. Turner}
\affiliation{Green Bank Observatory, P.O. Box 2, Green Bank, WV 24944, USA}

\author[0000-0001-8885-6388]{Timothy Dolch}
\affiliation{Department of Physics, Hillsdale College, 33 E. College Street, Hillsdale, MI 49242, USA}
\affiliation{Eureka Scientific, 2452 Delmer Street, Suite 100, Oakland, CA 94602-3017, USA }

\author[0000-0002-4049-1882]{James M. Cordes}
\affiliation{Cornell Center for Astrophysics and Planetary Science and Department of Astronomy, Cornell University, Ithaca, NY 14853, USA}

\author[0000-0002-4941-5333]{Stella K. Ocker}
\affiliation{Cahill Center for Astronomy and Astrophysics, California Institute of Technology, Pasadena, CA 91101, USA}
\affiliation{The Observatories of the Carnegie Institution for Science, Pasadena, CA 91101, USA}

\author[0000-0002-1797-3277]{Daniel R. Stinebring}
\affiliation{Department of Physics and Astronomy, Oberlin College, Oberlin, OH 44074, USA}

\author[0000-0002-2878-1502]{Shami Chatterjee}
\affiliation{Cornell Center for Astrophysics and Planetary Science and Department of Astronomy, Cornell University, Ithaca, NY 14853, USA}

\author[0000-0001-7697-7422]{Maura A. McLaughlin}
\affiliation{Department of Physics and Astronomy, West Virginia University, P.O. Box 6315, Morgantown, WV 26506, USA}
\affiliation{Center for Gravitational Waves and Cosmology, West Virginia University, Chestnut Ridge Research Building, Morgantown, WV 26505, USA}

\author[0000-0002-4925-8403]{Victoria E. Catlett}
\affiliation{Green Bank Observatory, P.O. Box 2, Green Bank, WV 24944, USA}

\author[0000-0002-4188-6827]{Cody Jessup}
\affiliation{Department of Physics, Hillsdale College, 33 E. College Street, Hillsdale, MI 49242, USA}
\affiliation{Department of Physics, Montana State University, Bozeman, MT 59717, USA}

\author[0009-0009-7809-3335]{Nathaniel Jones}
\affiliation{Department of Physics, Hillsdale College, 33 E. College Street, Hillsdale, MI 49242, USA}

\author[0009-0007-2434-2776]{Christopher Scheithauer}
\affiliation{Department of Physics, Hillsdale College, 33 E. College Street, Hillsdale, MI 49242, USA}

%% file: abstract.tex
\begin{abstract}
We use cyclic spectroscopy to perform high frequency-resolution analyses of multi-hour baseband Arecibo observations of the millisecond pulsar PSR B1937+21. This technique allows for the examination of scintillation features in far greater detail than is otherwise possible under most pulsar timing array observing setups. We measure scintillation bandwidths and timescales in each of eight subbands across a 200 MHz observing band in each observation. Through these measurements we obtain intra-epoch estimates of the frequency scalings for scintillation bandwidth and timescale.Thanks to our high frequency resolution and the narrow scintles of this pulsar, we resolve scintillation arcs in the secondary spectra due to the increased Nyquist limit, which would not have been resolved at the same observing frequency with a traditional filterbank spectrum using NANOGrav's current time and frequency resolutions, and the frequency-dependent evolution of scintillation arc features within individual observations. We observe the dimming of prominent arc features at higher frequencies, possibly due to a combination of decreasing flux density and the frequency dependence of the plasma refractive index of the interstellar medium. We also find  agreement with arc curvature frequency dependence predicted by \citet{OG_arcs} in some epochs. Thanks to the frequency resolution improvement provided by cyclic spectroscopy, these results show strong promise for future such analyses with millisecond pulsars, particularly for pulsar timing arrays, where such techniques can allow for detailed studies of the interstellar medium in highly scattered pulsars without sacrificing the timing resolution that is crucial to their gravitational wave detection efforts. 
\end{abstract}